\documentclass[aps,epsf,preprint]{revtex4}
\usepackage{amssymb}
\usepackage{color}

\def\1{{\bf 1}}
\def\[{\left[}
\def\]{\right]}
\def\be{\begin{eqnarray}}
\def\ee{\end{eqnarray}}
\def\bm{\begin{pmatrix}}
\def\em{\end{pmatrix}}
\def\nn{\nonumber}
\def\({\left(}
\def\){\right)}
\def\bk#1{\langle#1\rangle}
\def\eq#1{(\ref{#1})}

\def\a{\alpha}
\def\s{\sigma}

\def\e{\epsilon}

\def\f{\phi}

\def\l{\lambda}
\def\m{\mu}
\def\x{\times}
\def\ket#1{|#1\rangle}

\def\d{\delta}

\def\labels#1{\label{#1}}
\def\edc{\end{document}}

\def\Rw{\Rightarrow}
\def\bn{\begin{enumerate}}
\def\i{\item}
\def\en{\end{enumerate}}
\def\b{\beta}
\def\g{\gamma}

\def\ba{\begin{array}}
\def\ea{\end{array}}
\def\bc{\begin{center}}
\def\ec{\end{center}}

\def\edoc{\end{document}}

\def\^{$\wedge$}
\def\.{\!\cdot\!}

\def\igw#1{\includegraphics[width=#1cm]}
\def\+{\!+\!}
\def\-{\!-\!}

\def\bsl{\backslash}

\def\sb#1{\!\stackrel{.}{#1}\!}

\def\1{\sb 1}
\def\2{\sb 2}
\def\3{\sb 3}
\def\4{\sb 4}
\def\5{\sb 5}
\def\6{\sb 6}
\def\7{\sb 7}
\def\8{\sb 8}
\def\9{\sb 9}
\def\M{M\"obius\ }
\usepackage{graphicx}
\graphicspath{/Users/harrylam/Current/filetex/Helicity/graphics/}
\def\pf{{\rm Pf}}
\def\Q{\Psi}
\def\Tr{{\rm Tr}}
\def\n{\nu}

\def\=#1{\stackrel{#1}{=}}
\def\bb#1#2#3#4{{\bk{#2#3}[#1#4]\over\bk{#1#3}[#2#4]}}
\def\cp#1#2#3{{[#2#1]\bk{#2#3}\over\bk{#1#3}}}  
\def\cm#1#2#3{{\bk{#2#1}[#2#3]\over[#1#3]}}  
\def\h#1{{\hat#1}}
\def\S{Sec.~}
\def\h{{1\over 2}}

\begin{document}

\title{Evaluation of the CHY Gauge Amplitude}
\author{C.S. Lam$^a$  and York-Peng Yao$^b$}
\address{$^a$Department of Physics, McGill University\\
 Montreal, Q.C., Canada H3A 2T8\\
$^a$Department of Physics and Astronomy, University of British Columbia,  Vancouver, BC, Canada V6T 1Z1 \\
$^b$Department of Physics, The University of Michigan
Ann Arbor, MI 48109, USA\\
Emails: Lam@physics.mcgill.ca\  yyao@umich.edu}


\begin{abstract}
The Cachazo-He-Yuan (CHY) formula for $n$-gluon scattering is known to give the same
amplitude as the one obtained from  Feynman diagrams, though the former contains neither vertices nor propagators
explicitly. The equivalence was shown by indirect means, not by a direct evaluation of the $(n\! - \!3)$-dimensional integral in the
CHY formula. The purpose of this paper is to discuss how such a direct evaluation can be carried out. 
There are two basic difficulties in the calculation: how to handle the large number of terms
in the reduced Pfaffian, and how to carry out the integrations in the presence of a $\sigma$-dependence much more
complicated than the Parke-Taylor form found in a CHY double-color scalar amplitude. We have solved both
of these problems, and have formulated a method  that can be applied to any $n$. 
Many examples are provided to illustrate these calculations. 

\end{abstract}
\narrowtext
\maketitle

\section{Introduction}

Cachazo, He, and Yuan (CHY) proposed a formula for the $n$-gluon scattering amplitude in the tree approximation \cite{CHY1,CHY2,CHY3,CHY4}. 
Unlike the textbook amplitude which is made up of a sum of Feynman diagrams, each  built from a collection of vertices and propagators, 
the CHY amplitude is a single global formula given by an $(n\-3)$-dimensional complex integral. See equation \eq{mg} 
in \S II 
for the exact expression. It is not immediately obvious how to break it down into a sum of Feynman amplitudes, much less to obtain
the local structure of vertices and propagators within each Feynman diagram. 
To simplify description,  `Feynman amplitudes' will be used throughout this paper to stand for amplitudes obtained
in the text-book manner by summing Feynman diagrams with Feynman rules.
Yet, it is known since the beginning \cite{CHY1,CHY2,CHY3,CHY4}
that it factorizes correctly in the collinear and the soft-gluon limits, just like Feynman tree amplitudes do, and its equivalence to the
Feynman amplitude was proven \cite{DG1} by showing that it  satisfies the Britto-Cachazo-Feng-Witten (BCFW) on-shell factorization formula \cite{BCFW}. Like the Feynman scattering amplitude, it is gauge invariant in the sense that the same result is obtained
by choosing any set of gauge-equivalent external polarization vectors. But unlike the Feynman amplitude, there is no need to fix a gauge for
the propagators, simply
because there are no propagators in the CHY formalism. Instead, a new kind of `gauge' emerges from the \M invariance
of the CHY amplitude. For concrete computations, it is necessary to choose three \M constants $\s_r, \s_s, \s_t$, as well as the Pfaffian lines
$\l$ and $\n$ in the CHY formula to be discussed in the next section, in spite of the fact that
 the final scattering amplitude is independent of these choices. In this respect
it is similar to the necessity of choosing a gauge for the propagators in evaluating a Feynman amplitude.

Although we know that the CHY amplitude must be the same as the Feynman amplitude, it is not easy to calculate the amplitude
directly from the CHY formula. $(n\-3)$ complex integrations must be performed, from which the momentum poles given
by the Feynman propagators must emerge. It must also give the Feynman-amplitude numerators without the benefit of 
 triple-gluon and four-gluon vertices which are absent in the CHY formalism. 
 The purpose of this paper is to discuss how the calculation can be carried out for
any $n$. 
It turns out that the CHY amplitude arranges its terms quite differently than the Feynman amplitude.
 Whereas the latter is grouped according to Feynman diagrams with a fixed set of propagators in the denominators,
 the former is grouped according to a fixed pattern of its numerators. The numerators of a Feynman amplitude, assembled
 from the vertices, have a complicated form after a straight forward expansion. It depends on the topology of the Feynman diagram and 
 seems to have no discernible
 pattern between diagrams. Its expansion into a function of  dot products of external momenta and polarization vectors 
 can only be obtained through tedious algebra. 
 The denominators of a CHY amplitude corresponding to a fixed pattern of the numerator has to be calculated
 by integration from its specific $\s$-dependence, but a set of rules can be
 developed to determine the associated propagators.

Before going into details of how computations are to be carried out, it might be useful to have a rough idea
of the role played by each part of the CHY integrand.  Other than the integration measure and 
a normalization factor, the integrand  consists of two main parts. 
The first contains
the product of scattering functions $f_i$, present universally in all the CHY amplitudes. It
is the source of all propagators in the amplitude. Every propagator of every Feynman diagram
originates here, so it is like a stem cell before differentiation. 
The differentiation or selection control comes from the 
$\sigma$-dependent factors in the rest of the integrand.
It picks out one or several dominant regions of integration, 
from which one or several Feynman diagrams emerge. 
The second part consisting of the reduced Pfaffian $\pf'(\Q)$ and a Parke-Taylor factor is where these
controlling $\sigma$ factors lie. The reduced Pfaffian consists of 
many terms having different polarizations and momenta, each with a different $\s$ dependence, and
it is possible to group together terms in the reduced Pfaffian with the same $\s$ dependence.
As a result, the CHY amplitudes can be arranged according to
numerator factors coming from terms of the reduced Pfaffians
with the same $\s$ dependence, rather than identical denominator factors like in a Feynman amplitude.
Integration over its $\s$ dependence then produces products of propagators corresponding to one or several Feynman diagrams.

The groupings of the reduced Pfaffian with the same $\s$ factors will be discussed in \S III. 
There is a very useful `shift invariance' for the reduced Pfaffian which can be used to check
calculations. That will be discussed in Sec.~IV.

As in the Feynman amplitude evaluation, simplification can be obtained by choosing a suitable representation
for the transverse polarization vectors of the gluons. Polarization vectors used in the spinor-helicity
technique offers great simplifications, resulting for example in the celebrated Parke-Taylor formula \cite{PT} for $n$-gluon 
amplitudes when all but two legs carry the same helicity. The same choice also simplifies CHY amplitude calculations. 
For easy reference, we will refer to this choice of polarization vectors as the `helicity gauge'. Its general properties 
are reviewed and how it is useful in calculating 
the CHY amplitude is explained in \S V. 

\S VI discusses how the complex integrations in \eq{mg} can be carried out using a technique developed in a previous paper \cite{LYscalar}
for the CHY double-color scalar amplitude. Dominant regions in the $(n\-3)$-dimensional complex space are picked out
to present a pole in each of the $(n\-3)$ successive integrations, to enable residue calculus to be used for evaluating the integral.  
This technique is  applied to the evaluation of the $n=3,\ n=4$, and $n=5$ amplitudes respectively in Secs.~VII, VIII, and IX,
and to the evaluation of amplitudes of a general $n$ in \S X. Double poles are discussed in \S XII.

The expansion of the reduced Pfaffian depends on the choice of $\l$ and $\n$, and the selection
of the dominant integration regions depends on $r,s,t$, though not on the values $\s_r, \s_s, \s_t$. 
This freedom is illustrated in the explicit examples in Secs.~VII and VIII. Calculation details differ from one choice to another,
so with experience one can exploit this freedom to make a choice best suited for the problem. This independence can also be
used to check the result of the calculations, as is done in Secs.~VII and VIII.

In the CHY approach, the triple-gluon vertex of the Feynman amplitude can be reproduced in a $n=3$ calculation, as is done in
\S VII. To get the four-gluon vertex, one must carry out an $n=4$ calculation, without choosing the helicity gauge or any other gauge, 
then subtract out the contributions coming from the triple-gluon vertices. This 
is carried out in Sec.~XI. The calculation is somewhat
lengthy, but it does explicitly demonstrate the
equivalence between the $n=4$ CHY amplitude and the Feynman amplitude in any gauge. 

Finally a summary is provided in \S XIII.

\section{CHY Amplitude}
A color-stripped $n$-gluon scattering amplitude is given by the CHY formula \cite{CHY2} to be
\be
M^{\a}=\(-{1\over 2\pi i}\)^{n-3}\oint_\Gamma\s_{(rst)}^2\(\prod_{i=1,i\not=r,s,t}^n{d\s_i\over f_i}\){\pf'\Q\over\s_{(\a)}},\labels{mg}\ee
where $\a=(\a_1\a_2\cdots\a_n)$ describes the color, expressed as a permutation of $S_n$, in cycle notation. The scattering functions and the
$\s$-factors are
\be f_i&=&\sum_{j=1,j\not=i}^n{2k_i\.k_j\over \s_{ij}},\quad (1\le i\le n)\nn\\
\s_{ij}&=&\s_i-\s_j,\quad \s_{(rst)}=\s_{rs}\s_{st}\s_{tr},\nn\\
\s_{(\a)}&=&\s_{(\a_1\a_2\cdots\a_n)}=\prod_{i=1}^n\s_{\a_i\a_{i+1}}=\s_{[\a_1\a_2\cdots\a_n]}\s_{\a_n\a_1},\quad n+1\equiv 1.\labels{f}\ee
The three lines $r,s,t$ for the \M constants $\s_r,\s_s,\s_t$ will be referred to as  constant lines,
the rest variable lines.
The reduced Pfaffiann $\pf'\Q$
is invariant under any permutation of the external particles. It
is related to the Pfaffian of a matrix $\Q^{\l\n}_{\l\n}$ by
\be \pf'\Q={(-1)^{\l+\nu+\h n(n+1)}2^{n-3}\over\s_{\n\l}}\pf\(\Q^{\l\n}_{\l\n}\),\quad (\l<\n),\labels{ln}\ee
where $\Q^{\l\n}_{\l\n}$ is obtained from
 the matrix $\Q$ with its $\l$th and $\n$th columns and rows removed. 
We need this normalization of $\pf'\Q$, rather than the one used in Ref.~\cite{CHY2},
to reproduce the Parke-Taylor formula \cite{PT}. Different normalizations may be more convenient for other purposes.
The antisymmetric matrix $\Q$ is made up of three $n\x n$ matrices $A, B, C$,
\be
\Q=\bm{A&-C^T\cr C&B\cr}\em.\labels{QABCD}\ee
The non-diagonal elements of these three sub-matrices are
\be A_{ij}={k_i\.k_j\over\s_{ij}},\quad B_{ij}={\e_i\.\e_j\over\s_{ij}},\quad C_{ij}={\e_i\.k_j\over\s_{ij}},\quad 
-C^T_{ij}={k_i\.\e_j\over\s_{ij}},\quad 
(1\le i\not=j\le n),\labels{ABC}\ee
where $\e_i$ is the polarization of the $i$th gluon, satisfying $\e_i\.k_i=0$. 
The diagonal elements of $A$ and $B$ are zero, and that of $C$ is defined by
\be C_{ii}=-\sum_{j=1}^nC_{ij},\labels{CC}\ee
so that the column and row sums of $C$ is zero. A similar property is true for $A$ if the scattering equations $f_i=0$
are obeyed. This is the case because the integration contour $\Gamma$ encloses these zeros anticlockwise. 

For massless  particles satisfying momentum conservation, the amplitude $M^\a$ is independent on the choice of
$\l$ and $\n$. It is also gauge invariant, in the sense that when any $\e_i$ is replaced by $k_i$, then the amplitude is zero.

If $i$ is a number between 1 and $2n$, its {\it complement} $i'$ will be defined to be $i-n$ if $i>n$, and  $i+n$ if $i\le n$.
One of $i,i'$ is between 1 and $n$, and that number will be denoted by $\bar i$. With this notation, the matrix elements
of $\Q$ for $\bar i\not=\bar j$ can be summarized in one line as
\be
\Q_{ij}={v_i\.v_j\over\s_{\bar i\bar j}}, \quad v_{\bar i}=k_{\bar i},\ v_{\bar i+n}=\e_{\bar i}, \quad\mbox{if $\bar i\not=\bar j$}.\labels{matele}\ee
For $\bar i=\bar j$, we have
\be \Q_{\bar i\bar i}=\Q_{\bar i+n,\bar i+n}=0,\quad \Q_{\bar i+n,\bar i}=C_{\bar i\bar i}=-\Q_{\bar i,\bar i+n}.
\labels{mateld}\ee

The momenta $k_\l$ and $k_\n$ do not seem to appear in 
$\pf'\Q$ because the rows and columns containing them are absent in $\Q^{\l\n}_{\l\n}$.  Yet, the numerator of a Feynman amplitude generally depends on all momenta, so how can these two momenta be absent?
The answer is, they are there but hidden inside $C_{ii}$. Recall that $C_{ii}=-\sum_{j\not=i}\e_i\.k_j/\s_{ij}$, and the sum includes $j=\l$ and $\nu$.
Whereas every other matrix element in $\Q$ consists of only one term, with simple and regular dependence on $\e, k$, and $\s$, 
the element $C_{ii}$
is given by a sum so it has a complicated dependence on these variables. This fact complicate the evaluation of the amplitude as we shall
see later.

\section{Pfaffian Decomposition}

Let $\Q$ be a $2n\x 2n$ antisymmetric matrix, and $p=[p_1p_2\cdots p_{2n}]\in S_{2n}$ be a permutation of the $2n$ numbers
$[12\cdots 2n]$, with signature $(-)^p$. Let 
\be
\Q_p=(-)^p\prod_{\ell=1}^n\Q_{p_{2\ell-1}p_{2\ell}}.\labels{y}\ee
Two permutations which differ by interchanging  their $(2\ell-1)$th and $(2\ell)$th
elements give the same $\Q_p$, because $\Q$ is antisymmetric, and two permutations which permute the factors of \eq{y} 
also yield the same $\Q_p$.
In what follows we would use $\sum'$ to indicate the sum over those independent
permutations in $S_{2n}$ which give rise to distinct $\Q_p$. Then a Pfaffian of $\Q$ is defined by
\be
\pf(\Q)={\sum_p}'\Q_p={1\over 2^n n!}\sum_{p\in S_{2n}}\Q_p.\ee
The number of terms in $\pf(\Q)$ is $(2n)!/2^nn!=(2n-1)!!$.

The denominators of the non-diagonal matrix elements of $A,B,C$ in \eq{ABC} are the same. That property allows a simplification of $\pf(\Q)$
in \eq{QABCD}  from a sum over $p\in S_{2n}$ to  a sum over $\bar p\in S_n$. To do so,
take the term $\Q_p$ in \eq{y} and arrange the order of its factors $\Q_{p_{2\ell-1}p_{2\ell}}$ so that 
every $\Q_{pq}$ is followed by a
$\Q_{q'r}$, where $q'$ is the complement of $q$ defined at the end of the last section. 
That may require the use of antisymmetry  to convert $\Q_{rq'}$ to $-\Q_{q'r}$, but even so, this arrangement cannot 
continue if $q'$ has already been used up, because every number between 1 and $2n$  can occur only once.
In that case start over
again with any other number not already used. In that way, up to a sign, $\Q_p$ can finally be written in the form
\be
\Q_p=\pm\(\Q_{i_1'i_2}\Q_{i_2'i_3}\cdots\Q_{i_x'i_1}\)\(\Q_{j_1'j_2}\Q_{j_2'j_3}\cdots\Q_{j_y'j_1}\)\cdots
\(\Q_{k_1'k_2}\Q_{k_2'k_3}\cdots\Q_{k_z'k_1}\).\labels{psip}\ee
If we interchange $i_a$ and $i'_a$ and sum up the two terms, for $2\le a\le x$, then the first factor in \eq{psip} can be 
written  as 
\be k_{\bar i_1}\.U_{\bar i_2}\cdots U_{\bar i_x}\.\e_{\bar i_1}&&,\quad {\rm if}\ i_1'=\bar i_1,\nn\\
-\e_{\bar i_1}\.U_{\bar i_2}\cdots U_{\bar i_x}\.k_{\bar i_1}&&,\quad {\rm if}\ i_1'=(\bar i_1)',\labels{melt}\ee
where $U$ is a dyadic with respect to Lorentz indices, defined by
 \be
U_{\bar i}=k_{\bar i}\e_{\bar i}-\e_{\bar i}k_{\bar i}.\labels{V}\ee
Summing also over $i_1'=\bar i_1$ and $\bar i_1'$, the first factor of \eq{psip} becomes $\Tr(U_{\bar i_1}U_{\bar i_2}\cdots U_{\bar i_x})$.
Doing the same thing for the other factors of \eq{psip} would result in an expression for
$\pf(\Q)$ in which the sum over $p\in S_{2n}$ is reduced to a sum over $\bar p\in S_n$:
\be
\pf{\Q}=(-1)^{{1\over 2}n(n+1)}\sum_{\bar p\in S_n}(-)^{\bar p}\Q_{\bar p}=(-1)^{{1\over 2}n(n+1)}\sum_{\bar p\in S_n}(-)^{\bar p}\Q_I\Q_J\cdots\Q_K,\labels{pfsign}\ee
where
\be
 {I}=(\bar i_1\bar i_2\bar i_3\cdots\bar i_x),\quad 
{J}=(\bar j_1\bar j_2\bar j_3\cdots \bar j_y),\quad \cdots\quad
 {K}=(\bar k_1\bar k_2\bar k_3\cdots \bar k_z) \labels{pijk}\ee
 are the cycles of the permutation $\bar p\in S_n$. The cycle factors are given by
 \be
 \Q_I={U_I\over\s_I}={\h\Tr(U_{\bar i_1}U_{\bar i_2}\cdots U_{\bar i_x})\over \s_{\bar i_1\bar i_2}\s_{\bar i_2\bar i_3}\cdots\s_{\bar i_x\bar i_1}}\labels{cycle}\ee
 when the cycle length $x>1$, and
for a 1-cycle, the cycle factor is
 \be
 \Q_{(\bar i)}=C_{\bar i\bar i}=-\sum_{\bar j\not=\bar i}{c_{\bar i\bar j}\over\s_{\bar i\bar j}}.\ee
Similar expressions apply to all the other cycles including $J$ and $K$.

The factor $\h$ in \eq{cycle} comes about because of  double counting. For $x>1$, the cycles $(\bar i_1\bar i_2\cdots \bar i_x)$
and $(\bar i_x\cdots \bar i_2\bar i_1)$ have identical cycle factors, so if we sum up all permutations $\bar p\in S_n$ in 
\eq{pfsign}, a factor $\h$ is called for.

To illustrate the formulas and notations, suppose $n=4$. Consider the permutation $p=[15247368]$ 
of [12345678] whose signature is $(-)^p=+1$. Its matrix element is
\be\Q_p&=& \Q_{15} \Q_{24} \Q_{73} \Q_{68}=\Q_{51} \Q_{24} \Q_{86}\Q_{73} \nn\\
&=& C_{11}{(k_2\.k_4)(\e_4\.\e_2)\over\s_{(24)}}C_{33},\quad \s_{(24)}=\s_{24}\s_{42}.\labels{psi4a}\ee
The associated $\bar p\in S_4$ is
$\bar p= (1) (24)(3)$, with $ I=(1), J=(24)$, $K=(3)$ and signature $(-)^{\bar p}=-1$. The associated cycle factor
 \be (-)^{\bar p}\Q_{\bar p}&=&-\Q_I\Q_J\Q_K= -C_{11}{\h\Tr[U_2U_4]\over\s_{(24)}}C_{33}
 =-C_{11}{\h\Tr[(k_2\e_2-\e_2k_2)(k_4\e_4-\e_4k_4)]\over\s_{(24)}}C_{33}\nn\\
&=&-C_{11}{\[(\e_2\.k_4)(\e_4\.k_2)-(\e_2\.\e_4)(k_2\.k_4)\]\over\s_{(24)}}C_{33} \ee
does include \eq{psi4a} as it should.

The Pfaffian $\pf(\Q)$ for $n=4$ has $7!!=105$ terms, which  can be grouped into a sum over the cycles of $S_4$. These cycles are 
\be {\cal C}=\{&&(1)(234), (1)(243), (2)(134), (2)(143), (3)(124), (3)(142), (4)(123), (4)(132),\nn\\
&& (1)(2)(34), (1)(3)(24), (1)(4)(23), (2)(3)(14), (2)(4)(13), (3)(4)(12), (1)(2)(3)(4),\nn\\
&&(12)(34), (13)(24), (14)(23), (1234), (1243), (1324), (1342), (1423), (1432)\quad \}.\labels{cycleC}\ee

Back to the general discussion. What we really need for the scattering amplitude is not $\pf(\Q)$, but
 $\pf(\Q^{\l\n}_{\l\n})$, where $\Q^{\l\n}_{\l\n}$  is $\Q$ with
the $\l$th and $\n$th columns and rows removed,  $1\!\le\! \l<\!\n\le\! n$. Note that although $\l$ and $\n$ are absent,
the columns and rows $\l', \n'$ are still present.
In the first group of terms in \eq{psip}, suppose we let $i_1'=\l'$. Since $\l,\n$ are not present in 
$\Q^{\l\n}_{\l\n}$, the
arrangement in this group could only end with a factor of the form $\Q_{i_x\n'}$. This is the only change
for $\pf(\Q^{\l\n}_{\l\n})$, the other groupings
$J,\cdots,K$ remain the same as in $\pf(\Q)$. Since an extra factor $1/\s_{\n\l}$ is present in
 $\pf'\Q=(-1)^{\l+\n+\h n(n+1)}2^{n-3}\pf(\Q^{\l\n}_{\l\n})/\s_{\n\l}$,
as long as we stipulate that the cycle $I$ must be of the form $(\l\bar i_2\cdots \bar i_{x-1}\n)$, the $\s$ factors in  $\pf'\Q$
are identical to those in $\pf\Q$. The trace in  $U_I$ is replaced by the matrix element
\be W_I=\e_\l.\(U_{i_2}U_{i_3}\cdots U_{i_{x-1}}\)\.\e_\n,\labels{W}\ee
and everything else remains essentially the same. In this way we get
\be
\pf'\Q=-2^{n-3}{\sum_{p\in S_n}}'(-)^p{W_IU_J\cdots U_K\over\s_{p}},\quad \s_p=\s_I\s_J\cdots \s_K,\labels{pfprime}\ee
where the prime on the summation sign indicates that the sum is taken over all $p\in S_n$ such that $\nu$ is changed into $\l$.
There are $(n-1)!$ such permutations in $S_n$ so the sum consists of $(n-1)!$ terms.
Note that we have dropped the bar on top of $p$ because from now on we always talk about cycles in $S_n$, never $S_{2n}$.

We shall refer to the cycle bounded at two ends by $\l$ and $\n$ as an {\it open cycle}, and the rest as {\it closed cycles}.
Open cycles are denoted by a square bracket, and closed cycles are denoted
by a round bracket. Thus $I=[\l i_2i_3\cdots i_{x-1}\n]$, but $J=(j_1j_2\cdots j_y)$.
Every matrix element in a closed cycle has the same number of $\e$ and $k$, hence the same number of $\e\.\e$
and $k\.k$, whereas the matrix element in the open cycle has two more $\e$ than $k$, hence there is one more
$\e\.\e$ than $k\.k$. In particular, there must be at least one $\e\.\e$ in this matrix element. This property will be
crucial in rendering many cycle factors zero in the `helicity gauge'.

There is another simple thing to note. Every number $a$ between 1 and $n$ must appear in every $p\in S_n$ once.
If it resides in $U_a$, then both $\e_a$ and $k_a$ must appear. However, if $a=\l$ or $\n$, then only $\e_a$
is present, not $k_a$.

Gauge invariance is the statement that $\pf'\Q=0$ if an $\e_i$ is replaced by $k_i$. This is easy to see in the present
formalism because such replacement renders $U_i=\e_ik_i-k_i\e_i$ zero.

The gauge amplitude \eq{mg} is invariant under a \M transformation, $\s_i\to(\a\s_i+\b)/(\g\s_i+\d), \a\d-\b\g=1$.
This is so because under such a transformation,
\be \s_{ij}&\to&\s_{ij}/(\g\s_i+\d)(\g\s_j+\d),\nn\\ 
C_{ii}&\to&C_{ii}(\g\s_i+\d)^2,\nn\\
\s_{(\a)}&\to&L\s_{(\a)},\nn\\
\pf'\Q&\to&\pf'\Q/L,\nn\\
\(\s_{(rst)}^2\prod_{a=1,a\not=r,s,t}^n{d\s_a\over f_a}\)&\to& L^2\( \s_{(rst)}^2\prod_{a=1,a\not=r,s,t}^n{d\s_a\over f_a}\),\nn\\
L&\equiv&1/\prod_{j=1}^n(\g\s_j+\d)^2.\ee
It is this invariance of \eq{mg} that allows $\s_r,\s_s,\s_t$  to be chosen freely. 
Now it can be verified that every one of the $(n-1)!$ terms in \eq{pfprime} also transforms
like $\pf'\Q$, hence when \eq{pfprime} is used to expand the amplitude, every term of the amplitude is also
\M invariant. This allows the values of $\s_r,\s_s,\s_t$ to be chosen differently for different terms, 
a freedom which may be useful in simplifying calculation.

Let us illustrate the difference between the cycles  in $\pf(\Q)$ and $\pf'(\Q)$ with the $n=4$ example, 
assuming $\l=1$ and $\n=2$. The allowed cycles 
for $\pf'\Q$ are now reduced to
\be\{[142](3),  [132](4), [12](3)(4), [12](34), [1342], [1432]\},\labels{cycle12}\ee
a total of $3!=6$ terms, far smaller than the number of terms of $\pf(\Q)$ appearing in \eq{cycleC}.

It is convenient to give the factors encountered in \eq{pfprime} names to make them simpler to write.
Accordingly, the following abbreviations will be used throughout this article:
\be
a_{ij}=a_{ji}=k_i\.k_j,\quad b_{ij}=b_{ji}=\e_i\.\e_j,\quad c_{ij}=\e_i\.k_j.\labels{name}\ee
We may also add a superscript to indicate the helicity of a polarization vector $\e$. For example,
$b_{ij}^{+-}=\e_i^+\.\e_j^-$.

\section{Shift invariance}
Add a multiple of the $(i\+n)$th column of $\Q$ to the $i$th column, and the same multiple
of the $(i\+n)$th row to the $i$th row. This action changes the matrix $\Q$ to a `shifted matrix' $\hat\Q$
by replacing $k_i$ with the shifted momentum $\hat k_i=k_i+z_i\e_i$, 
where  $z_i$ is an arbitrary complex constant. 
This action does not change the Pfaffian,  $\pf\Q=\pf\hat\Q$, nor the reduced Pfaffians, 
$\pf\Q^{\l\n}_{\l\n}=\pf\hat \Q^{\l\n}_{\l\n}$, provided $i\not=\l,\n$. The latter condition is necessary because
$k_\l, k_\nu$ are not contained in $\pf'\Q$ except through $C_{aa}$, so a shift in either of them is meaningless.
This invariance, which is true in any gauge, will be referred to as the {\it shift invariance}.

Shift invariance is a useful consequence of the CHY theory because it can be used to check a calculation of $\pf'\Q$. 
Examples of such checks will be given in the next few sections.
The shift $k_i\to \hat k_i=k_i+\e_i$ leads to the shifts

\be
a_{ji}\to a_{ j \hat i}&=&a_{ji}+c_{ij},\quad a_{ij}\to a_{\hat i j}=a_{ij}+c_{ij},\nn\\
 c_{ji}\to c_{j \hat i}&=&c_{ji}+b_{ji},\quad c_{ij}\to c_{\hat i j}=c_{ij},\quad(i\not=\l,\n);\nn\\
 b_{ij}\to b_{\hat i j}&=&b_{ij},\quad b_{ji}\to b_{j\hat i}=b_{ji},\nn\\
 C_{ii}\to C_{\hat i\hat i}&=&C_{ii},\quad({\rm all}\ i).\labels{sing}\ee

Similarly, if we add the $ith$ column to the $(i\+n)$th column, and the
$i$th row to the $(i\+n)$th row, the reduced Pfaffian also remains invariant, provided $i\not=\l,\n$. In terms
of $a,b,c$, this shift invariance is
\be
 c_{ji}\to c_{j \hat i}&=&c_{ji},\quad c_{ij}\to c_{\hat i j}=c_{ij}+a_{ij},\nn\\
 b_{ij}\to b_{\hat i j}&=&b_{ij}+c_{ji},\quad b_{ji}\to b_{j\hat i}=b_{ji}+c_{ji},\quad(i\not=\l,\n);\nn\\
 a_{ji}\to a_{ j \hat i}&=&a_{ji},\quad a_{ij}\to a_{\hat i j}=a_{ij},\nn\\ 
 C_{ii}\to C_{\hat i\hat i}&=&C_{ii},\quad({\rm all}\ i).\labels{sing2}\ee

We shall refer to \eq{sing} as shift invariance of the first kind, and \eq{sing2} as shift invariance of the second kind.
Shift invariance of the second kind is the same as Yang-Mills gauge invariance.

Let $\d_i$ refer to the change when $k_i\to k_i+\e_i$, and $\d'$ refer to the change when $\e_i\to\e_i+k_i$, then
\be
\d_ia_{ji}&=&\d_i a_{ij}=c_{ij},\quad \d_ic_{ji}=b_{ji},\quad \d_ic_{ij}=\d_ib_{ij}=\d_ib_{ji}=0,\nn\\
\d'_ib_{ji}&=&\d_i b_{ij}=c_{ji},\quad \d'_ic_{ij}=a_{ij},\quad \d'_ic_{ji}=\d'_ia_{ij}=\d'_ia_{ji}=0,\quad(i\not=\l,\n);\nn\\
\d_iC_{ii}&=&\d_i C_{jj}=\d'_iC_{ii}=\d'_iC_{jj}=0,\quad({\rm all}\ i).\labels{singt}\ee

Shift invariance can also be seen from the fact that $U_i=k_i\e_i-\e_ik_i$ in \eq{V} is invariant
under $k_i\to k_i+\e_i$ and $\e_i\to \e_i+k_i$.

Terms of the reduced Pfaffian with different $\s$-dependence
must be separately  shift invariant.

If the Pfaffian is calculated in the helicity gauge where the constraints \eq{hg1}, \eq{hg2}, and \eq{hg3} are used,
then only those changes respecting these constraints are shift invariant. In other words, any shift that changes
any of the $b_{ij}$ or $c_{ij}$ in \eq{hg1}, \eq{hg2}, and \eq{hg3} should not be applied to $\pf'\Q$.

\section{Helicity Gauge}
The number of terms in $\pf'\Q$ for an $n$-particle amplitude is $(2n-3)!!$, which is  15 for $n=4$, and already 105 for $n=5$.
In an attempt to reduce the number of terms to make the calculation manageable, we adopt the 
{\it helicity gauge}, defined by having the polarization vectors used in the spinor-helicity technique. 
Every particle $i$ carrying a $+$ helicity is required to satisfy $\e_i^+\.k_+=0$,
and every particle $i$ carrying a $-$ helicity is required to satisfy $\e_i^-\.k_-=0$. The {\it reference momenta}
$k_\pm$ are the momentum
of a line labelled $\pm$ and carry helicity $\mp$. In  spinor-helicity notation, 
\be \e_i^+={\ket{+}[i|\over\bk{+i}},\quad \e_i^-=-{\ket{i}[-|\over[i-]},
\quad k_\pm=\ket{\pm}[\pm|,\quad k_i=\ket{i}[i|.\labels{epsilon}\ee
They lead to the dot products
\be \e_i^\pm\.\e_j^\pm&=&0,\quad b_{ij}^{+-}:=\e_i^+\.\e_j^-={\bk{j+}[i-]\over\bk{i+}[j-]},\quad
a_{ij}:=k_i\.k_j=\bk{ij}[ji],\nn\\
c^+_{ij}&:=&\e_i^+\.k_j=-[ij]{\bk{j+}\over\bk{i+}},\quad c^-_{ij}:=\e_i^-\.k_j=-\bk{ij}{[j-]\over[i-]}.\labels{eprel}\ee
For simplicity in writing, the superscripts of $b$ and $c$ are often omitted.
From these relations, we see that many dot products are zero, which can be used to reduce the number of terms
in $\pf'\Q$. For instance,
\be 0&=&c_{i\pm}^\pm=\e_i^\pm\.k_\pm,\labels{hg1}\\
0&=&b_{+i}^{-+}=\e_+^-\.\e_i^+,\labels{hg2}\\
 0&=&b_{i-}^{-+}=\e^-_i\.\e^+_-,\labels{hg3}\\
 0&=&b^{+-}_{ij}a_{ji}-c^+_{ij}c^-_{ji}=(\e_i^+\.\e_j^-)(k_j\.k_i)-(\e_i^+\.k_j)(\e_j^-\.k_i)=-{1\over 2}\Tr(U_i^+U_j^-),\labels{hg4}\ee
where $U_a^\pm$ is \eq {V} with $\e_a$ replaced by $\e_a^\pm$,  and the subscripts $\pm$ stands
for the reference lines $j=\pm$. We also have
\be U_+U_-&=&(\e_+^-k_+-k_+\e_+^-)\.(\e_-^+k_--k_-\e_-^+)={1\over\bk{+-}[-+]}(\ell k_+-k_+\ell)\.(\ell k_--k_-\ell)\nn\\
&=& -\ell\ell=U_-U_+.\labels{upum}\ee

In terms of the complex light-like momentum 
\be \ell=\ket{+}[-|=\e_+^-[-+]=\e_-^+\bk{+-},\labels{ell}\ee
 \eq{epsilon} also implies
\be \ell\.\ell=\e_+^-\.\e_-^+=k_\pm\.\ell=\e_a^\pm\.\ell=0\quad(\forall a).\ee
Moreover, since $\e_+=\e_-{\bk{+-}\over [-+]}:=\e_-J$, we also have
\be
c_{+j}=Jc_{-j}.\labels{cpm}\ee

Recall that the factors of $\pf'\Q$  in the open cycle in \eq{pfprime} contains at least one $\e\.\e$. Thus
\bn
\i {\it the amplitude vanishes if every particle has the same helicity}, because then all $\e_i\.\e_j=0$;
\i{\it the amplitude also vanishes if all but one particle $a$ have the same helicity}. \quad
If we take $a$ to be the reference line for the majority helicity, then all $\e_i\.\e_j=0$, hence the amplitude vanishes.
\en

\section{Evaluation of the amplitude}
The gauge amplitude  $M^\a$ in \eq{mg} differs from the double-color scalar amplitude $m(\a|\b)$
 in only one aspect: the dynamical factor $1/\s_{(\a)}\s_{(\b)}$ in $m(\a|\b)$ is replaced by the dynamical
factor $\pf'\Q/\s_{(\a)}$ in $M^\a$. Nevertheless, this simple replacement makes the gauge amplitude much more
difficult to evaluate. To start with, $\pf'\Q$ contains $(2n-3)!!$ terms compared to the single term in $1/\s_{(\b)}$. 
This can be greatly simplified in the helicity gauge, for example,
\bn
\i[\label(i).]
\eq{hg2} implies that the open cycle must carry both helicities because it must contains at least one non-zero $\e\.\e$ factor;
\i[\label(ii).] \eq{hg3} implies that if $\l$ or $\n$ is $+$ or $-$, then the $W$-factor in the open cycle (see \eq{W}) must contain
at least one $U_i$;
\i[\label(iii).] \eq{hg4} implies that a closed 2-cycle must have the same helicity.
\en
The following relation of momentum conservation is also very useful in calculations,
\bn
\i[\label(iv).] \be\sum_{j=1}^nc_{ij}^\pm=\sum_{j\not=i}c_{ij}^\pm=0,\labels{mcn}\ee
\en
especially when coupled with equation \eq{hg1}.

Besides the large number of terms in $\pf'\Q$,
the other problem is  the complicated $\s$-dependence of $\pf'\Q$.  As shown in \eq{cycle} and \eq{pfprime}, 
it is given by many cycles of different lengths, unlike the double-color $\s_{(\b)}$ which only consists
 of a single $n$-cycle for the permutation $\b\in S_n$. Among these cycles, the 1-cycles are hard to handle because they
 consist of many terms with different $\s$'s, and the higher cycles may lead to double poles making
 the residue difficult to calculate.
 All of these make the integrations of $M^\a$ far more difficult to evaluate than
 the integrations in $m(\a|\b)$.

One way to deal with the complicated cycles in $\pf'\Q$
is to algebraically convert them to a combination of simple $n$-cycles \cite{FG,BBBD}, so that
the  trivalent \cite{CHY3}, polygon \cite{BBBD2}, or pairing \cite{LYscalar} 
rules designed for double-color scalar amplitudes can be used to evaluate the gauge amplitude. 
We shall illustrate this method below for $n=3, 4$, and 5. However,
there is no guarantee that the procedure will succeed, at least not easily. 
In particular, the matrix element $C_{\bar i\bar i}$ for a 1-cycle $(\bar i)$ has a complicated $\s$-dependence which
makes it difficult to be so converted for $n>5$, for reasons to be explained later.

Another way is to calculate \eq{mg} directly as it is, using
the `multi-crystal' method developed in \cite{LYscalar} for the CHY double-color scalar amplitude. That method is simply
a systematic way to locate regions in the $(n\-3)$-dimensional complex space that dominate the integral,
then use residue calculus to evaluate the $(n\-3)$ integrals one at a time. 
It will be adapted
here to compute the CHY gauge amplitude. 
It works well if only simple poles are present, because residues are then easy to calculate.
The technique discussed in this section and in \S X are designed with that in mind.
Terms with double poles, if present, must be computed separately. Fortunately, by suitably choosing the
gauge constants $r,s,t,\l,\n,+,-$, double poles can often be avoided. This is for example the case for $n=3, 4, 5$, which will be dealt with in
Secs.~VII, VIII, IX. 
Otherwise double poles can still be handled. Details can be found in Sec.~XII. 

To make writings simple but without scarifying generality, we shall assume for convenience
$\a=(123\cdots n)$ from now on. Different $\a$'s can be obtained simply by substitution.

In the present context, a  {\it crystal}  is a set  $S=\{i\+1, i\+2,\cdots,i\+m, i\+m\+1\}$  
of consecutive lines (line additions are understood to be mod $n$) that contains one and only one `defect'.
A {\it defect} is a line $r$ without the corresponding scattering function $f_r$ present in the integrand of the remaining integrations.
Originally, in \eq{mg}, only the constant lines qualify as defects, but later  on new defects will appear from
old `triggers'.

To explain what a `trigger' is, it is necessary to know how crystals are related to integrations. Each crystal gives rise to one
integration, with the integration variable $s$ defined by a scaling change  $\s_{x,x+1}=s\s'_{x,x+1},\ (i\+1\le x\le i\+m)$.
To ensure the right number of new variables, a constraint has to be put on $\s'_{x,x+1}$, and we shall do so by 
picking a variable line $p$ in $S$ and demand $\s'_{pr}=1$. This line $p$ is the {\it trigger}, and the relation
$\s'_{pr}=1$ is referred to as the {\it triggering relation}. For a motivation of these names please see Ref.~\cite{LYscalar}.

The integration in $s$ consists of computing the residue at the $s=0$ pole. To see when a pole appears,
note that on the one hand, every $1/f_x $ for $x\in S$ scale like $s$, so $\prod_{x\in S, x\not=r}(d\s_x/f_x)\sim s^{2m-1}$, 
and on the other hand, $1/\s_\a$ scales like $1/s^m$,
so a pole at $s=0$ occurs only when $\pf'\Q\sim 1/s^m$ or a higher power. The most efficient way to produce a high power of $1/s$
in $\pf'\Q$ is to have all the lines in $S$ appear adjacently in some cycle of $\pf'\Q$, in which case $\pf'\Q\sim 1/s^m$ if the cycle also
contains some other lines, and $\pf'\Q\sim 1/s^{m+1}$ if the cycle contains no other lines. In particular, a 1-cycle $(x)=C_{xx}\sim 1/s$
when $x\in S$. Putting these together, we see that the integrand of \eq{mg} produces a simple pole at $s=0$ if the lines of the crystal
all appear adjacently in a cycle which contains also other lines, and it produces a double pole if the cycle contains no other line.

Double poles will be discussed in Sec.~XII. For the rest of this section we will 
assume the poles to be simple poles.

After the integration, $f_p$ turns into the inverse propagator $s_S=s_{i+1,i+2,\cdots,i+m+1}$ \cite{LYscalar}.
As a result, $f_p$ is removed from the integrand so $p$ becomes a defect in subsequent integrations.

Two crystals are said to be {\it compatible} if they do not intersect, or one is completely inside the other.
Originally there are only three  defects, coming from  the three constant lines $r,s,t$ in \eq{mg}. 
Since each crystal is allowed  only one defect, the external lines of the amplitude can be
divided into three non-intersecting crystals. One of these three crystals may not contain any variable line
in which case it can be ignored in all the integrations. Each integration brings along a new defect $p$ 
which allows more ways to create smaller compatible crystals. After all the integrations, 
$(n-3)$ propagators appear which can be represented by a Feynman diagram.
Variable lines are replaced by defects, so
using the triggering relations, every $\s_{ij}$ becomes either $\pm 1, \pm \s_{rs}, \pm\s_{st}$, or $\pm\s_{ts}$.
The numerator of the gauge amplitude is then essentially $\pf'(\Q)$ with these substitutions of $\s_{ij}$.

The collection of a set of compatible crystals is referred to as a {\it multi-crystal}. The CHY amplitude is 
given by a sum over all possible independent multi-crystal sets, or a {\it complete set of multi-crystals}.

Owing to \M symmetry,
the amplitude \eq{mg} contains a large number of invariance, making it
independent of the choice of the constant lines $r,s,t$, the \M constants $\s_r, \s_s, \s_t$, the
Pfaffian lines $\l$ and $\n$, and the reference lines $+,-$ in the helicity gauge.
Different choices give rise to different surviving cycle structures in $\pf'\Q$ and different crystals, making
the calculations different, though at the end the result must be independent of the choice. One might
attempt to exploit this freedom to pick a choice most convenient for the calculation, but before having
the confidence to do so and the knowledge to know what to pick, and what difference do they make,
some explicit examples should be carried out. 
This is done in the next two sections
for  the $n=3$ and $n=4$ CHY
amplitudes, to show how different choices cause their computations to differ, and how at the end they all agree  
with the known result given by the Parke-Taylor formula \cite{PT}. The amplitude for $n\ge 5$ is computed in \S X,
whose outcome also agrees with the Parke-Taylor formula.

To simplify notations, we shall use a superscript to indicate the helicity of a particle $i$ in an amplitude, and a
subscript(s), if present, to indicate whether it is line $\l, \n, +$, or $-$. As in \S III,
a square bracket $[\cdots]$ is used to indicate the open cycle, and a round bracket $(\cdots)$ is used to indicate closed
cycles. We will also use the Mandelstam invariants defined by
\be
s_{ij\cdots k}=(k_i+k_j+\cdots+ k_k)^2\labels{mandelstam}\ee
in the calculations.

\section{\boldmath$n=3$}

There is no integration in $n=3$, so the three-point amplitude is simply 
$M=\s^2_{(123)}\break \pf'(\Q)/\s_{(123)}=\s_{(123)}\pf'\Q$. 
This amplitude is supposed to be 
independent of the \M constants $\s_{1,2,3}$, nor the choice of the Pfaffian lines $\l,\n$ in \eq{pfprime}.
In what follows, explicit calculations of  $1_\l2_\n 3$, $1_\n 2 3_\l$ and  $12_\l 3_\n$ are carried out 
 to verify this independence.

\subsection{Three-point vertex}
Table 1 shows the three choices of $\l$ and $\n$, the permutation $p\in S_3$ used in the calculation, and the resulting amplitude $M$.
These three expressions are identical because $b_{ij}=b_{ji}=\e_i\cdot\e_j$, and because
$c_{12}+c_{13}=c_{21}+c_{23}=c_{31}+c_{32}=0$ on account of momentum conservation.

$$\ba{|c|c|c|}\hline
\l,\n&p={\tt permutation}&M\\ \hline
1_\l 2_\n 3&[132], [12](3)&b_{13}c_{23}-b_{32}c_{13}+b_{12}c_{31}\\
1_\l 23_\n&[123], [13](2)&-b_{12}c_{32}+b_{23}c_{12}-b_{13}c_{21}\\
12_\l 3_\n&[213], [23](1)&b_{21}c_{31}-b_{13}c_{21}+b_{23}c_{12}\\
\hline\ea$$
\bc Table 1. Three-point vertex\ec

Here is the calculation of the first row using \eq{pfprime}:
\be 
M&=&-W_{[132]}{\s_{(123)}\over\s_{(132)}}+W_{[12]}{\s_{(123)}\over\s_{(12)}}C_{33},\nn\\
{1\over\s_{(12)}}C_{33}&=&{1\over\s_{(12)}}\(-{c_{31}\over\s_{31}}-{c_{32}\over\s_{32}}\)={1\over\s_{12}\s_{21}}c_{31}{\s_{21}\over\s_{32}\s_{31}}=-{c_{31}\over\s_{(123)}},\ \Rw\nn\\
M&=&-W_{[132]}+c_{31}W_{[12]}=b_{13}c_{23}-c_{13}b_{32}+c_{31}b_{12}.\ee
$M$ in the second row can be obtained from $M$ of the first row with $2\leftrightarrow 3$, times $-1$, and $M$
of the third row can be obtained from that of the second row with $2\leftrightarrow 1$, times $-1$.
The factor $-1$ comes from the ratio  $\s_{(123)}/\s_{(132)}$ and $\s_{(123)}/\s_{(213)}$.

We can use \eq{singt} to check the shift invariance of the second column in Table 1:
\be
\d_3[132]&=&\d_3(b_{13}c_{23}-b_{32}c_{13})=b_{13}b_{23}-b_{32}b_{13}=0, \quad\d_3\([12](3)\)=0,\nn\\
 \d_2[123]&=&\d_2(-b_{12}c_{32}+b_{23}c_{12})=-b_{12}b_{32}+b_{23}b_{12}=0, \quad \d_2\([13](2)\)=0,\nn\\
 \d_1[213]&=&\d_1(b_{21}c_{31}-b_{13}c_{21})=b_{21}b_{31}-b_{13}b_{21}=0,\quad \d_1\([23](1)\)=0,\nn\\
 \d'_3[132]&=&\d'_3(b_{13}c_{23}-b_{32}c_{13})=c_{13}c_{23}-c_{23}c_{13}=0, \quad\d'_3\([12](3)\)=0,\nn\\
 \d'_2[123]&=&\d'_2(-b_{12}c_{32}+b_{23}c_{12})=-c_{12}c_{32}+c_{32}c_{12}=0, \quad \d'_2\([13](2)\)=0,\nn\\
 \d'_1[213]&=&\d'_1(b_{21}c_{31}-b_{13}c_{21})=c_{21}c_{31}-c_{31}c_{21}=0,\quad \d'_1\([23](1)\)=0,\nn\\
 \ee

The Feynman triple-gluon vertex with the color  stripped off is 
\be
g\[(k_1-k_2)_\g  g_{\a\b}+(k_2-k_3)_\a g_{\b\g}+(k_3-k_1)_\b g_{\g\a}\],\labels{tgv}\ee
where $\a,\b,\g$ are the Lorentz index for momentum vectors $k_1, k_2, k_3$. If we contract this vertex with
the polarization vectors $\e_1,\e_2,\e_3$, and use momentum conservation, we get
\be 2g\[(\e_3\.k_1)(\e_1\.\e_2)+(\e_1\.k_2)(\e_2\.\e_3)+(\e_2\.k_3)(\e_3\.\e_1)\]
=2g\[c_{31}b_{12}+c_{12}b_{23}+c_{23}b_{31}\]=2gM,\ee
where $M$ is given by the last column of Table 1. Hence the CHY three-point amplitude is the usual
triple-gluon vertex with the coupling constant $g=\h$.

\subsection{Spinor-helicity expression}
The three-point amplitude $M=b_{13}c_{23}-b_{32}c_{13}-b_{12}c_{32}$ will now be expressed in spinor-helicity
language for the two different helicity configurations, $1^+2^+3^-$, and $1^-2^-3^+$. Note that in this simple case
there is no need to specify what the reference lines $\pm$ are. They may very well be lines outside of 1,2,3.

\subsubsection{$1^+2^+3^-$}
In this case $b_{12}=0$, hence
\be
M&=&b_{13}c_{23}-b_{23}c_{13}=\bb{1}{3}{+}{-}\cp{2}{3}{+}-\bb{2}{3}{+}{-}\cp{1}{3}{+}\nn\\
&=&{\bk{3+}^2\over\bk{1+}\bk{2+}[3-]}\([1-][32]-[2-][31]\)=-{\bk{3+}^2[21][3-]\over\bk{1+}\bk{2+}[3-]},
\ee
where Schouten identity was used in the last step. Now using momentum conservation, $\bk{3+}[31]=-\bk{2+}[21]$
and $\bk{3+}[32]=-\bk{1+}[12]$, hence
\be
M={[12]^4\over[12][23][31]},\ee
which agrees with the usual spinor-helicity result.

\subsubsection{$1^-2^-3^+$}
In this case
\be
M&=&b_{13}c_{23}-b_{23}c_{13}=\bb{3}{1}{+}{-}\cm{2}{3}{-}-\bb{3}{2}{+}{-}\cm{1}{3}{-}\nn\\
&=&{[3-]^2\over\bk{3+}[1-][2-]}\(\bk{1+}\bk{32}-\bk{2+}\bk{31}\)={[3-]^2\bk{12}\bk{3+}\over\bk{3+}[1-][2-]}={\bk{12}^4\over\bk{12}\bk{23}\bk{31}},
\ee
which agrees with the usual spinor-helicity result.

\section{\boldmath$n=4$}
We shall compare  calculations of various choices of $\l,\n,+,-$ in the helicity configuration $1^+2^+3^-4^-$,
with the constant lines taken to be
$r,s,t=1,2,3$ and the color taken to be $\a=(1234)$. In that case the amplitude \eq{mg} is 
\be
M={-1\over 2\pi i}\oint_\Gamma{d\s_4\over f_4}{\s_{(123)}^2\pf'\Q\over\s_{(1234)}}.\labels{mgn4}\ee 

The possible crystals are $S=\{34\}$ and $S=\{41\}$. Terms in $\pf'\Q$ proportional to $1/\s_{34}$ gives rise to $1/s_{34}$, and terms
proportional to $1/\s_{41}$ gives rise to $1/s_{41}$.

\subsection{\boldmath $1^+_{\l}2^+3^-4^-_{\n}$}
The allowed permutations are
\be  [1234], [1324], [124](3), [134](2), [14](23), [14](2)(3).\labels{n4perp}\ee
Depending on the choice of the reference lines $\pm$, some of these permutations are zeros. In what follows we will
consider several assignments of $(-,+)$.

We will also check the shift invariance \eq{singt} for each of the three cases below. Since $\l=1$ and $\n=4$, we should
have $\d_2(\pf'\Q)=\d_3(\pf'\Q)=\d'_2(\pf'\Q)=\d'_3(\pf'\Q)=0$, provided these changes do not affect
\eq{hg1}, \eq{hg2}, and \eq{hg3}.

\subsubsection{\boldmath $1^+_{\l-}2^+3^-4^-_{\n+}$}
The only non-zero $b$ is $b_{23}^{+-}$, so the open cycle must contain both 2 and 3, which leaves
only [1234] and [1324] as the allowed permutations. In this case,

\be \h\pf'(\Q)&=&{W_{[1234]}\over\s_{(1234)}}+{W_{[1324]}\over\s_{(1324)}}=b_{23}^{+-}\({c^+_{12}c^-_{43}\over\s_{(1234)}}+{c^+_{13}c^-_{42}\over
\s_{(1324)}}\),\nn\\
c^+_{12}c^-_{43}&=&[12]\bk{43}{\bk{24}[31]\over\bk{14}[41]},\nn\\
c^+_{13}c^-_{42}&=&[13]\bk{42}{\bk{34}[21]\over\bk{14}[41]}=c^+_{12}c^-_{43},\nn\\
{1\over\s_{(1234)}}+{1\over\s_{(1324)}}&=&{1\over\s_{23}\s_{41}}\({1\over\s_{12}\s_{34}}-{1\over\s_{13}\s_{24}}\)=-{1\over\s_{(1243)}},\nn\\
\h\pf'\Q&=&-{b^{+-}_{23}c^+_{12}c^-_{43}\over \s_{(1243)}}.
\ee

In this gauge $0=c_{14}=c_{34}=c_{31}=c_{41}=b_{13}=b_{14}=b_{24}$. These conditions must not be
violated in checking shift invariance. This means that we should not apply shifts $\d'_i$ of the second kind.
The only thing left is $\d_2$ and $\d_3$.
Shift invariance is trivial under $\d_2$ because $b_{12}=0$, and is equally trivial under $\d_3$ because $b_{43}=0$.

The integral to be carried out is just that of a double-color scalar amplitude, with $\a=[1234]$ and
$\b=[1243]$. This integral gives rise to $1/2k_1\.k_2=1/2k_3\.k_4$, hence
\be
M=-{b^{+-}_{23}c^+_{12}c^-_{43}\over k_3\.k_4}=-{\bk{34}[21]\over\bk{24}[31]}{[12]\bk{43}\bk{24}[31]\over\bk{14}[41]\bk{34}[43]}={[12]^4\over
[12][23][34][41]},\labels{m41}
\ee
which is the Parke-Taylor formula. Momentum conservation has been used to convert $\bk{43}/\bk{14}$ to $[12]/[32]$.

\subsubsection{\boldmath $1^+_{\l}2^+_-3^-_+4^-_{\n}$}
The only non-zero $b$ is $b^{+-}_{14}$, thus the only allowed permutations are [14](23) and [14](2)(3). 
With $2=-$ and $3=+$,
$c_{32}=c_{42}=0=c_{13}=c_{23}$, resulting in (23)=0. Hence we are
left with only [14](2)(3), which gives
\be
\h\pf'\Q&=&{W_{[14]}\over\s_{(14)}}C_{22}C_{33},\labels{a1}\\
W_{[14]}&=&b_{14}^{+-},\nn\\
C_{22}&=&-{c_{21}\over\s_{21}}-{c_{24}\over\s_{24}}=c_{24}\({1\over\s_{21}}-{1\over\s_{24}}\)=c_{24}{\s_{14}\over\s_{21}\s_{24}},\nn\\
C_{33}&=&-{c_{31}\over\s_{31}}-{c_{34}\over\s_{34}}=c_{34}\({1\over\s_{31}}-{1\over\s_{34}}\)=c_{34}{\s_{14}\over\s_{31}\s_{34}},\nn\\
\h\pf'\Q&=&{b_{14}^{+-}c^+_{24}c^-_{34}\over\s_{(1243)}},\nn\\
M&=&-{b_{14}^{+-}c^+_{24}c^-_{34}\over k_3.k_4}=-\bb{1}{4}{3}{2}\cp{2}{4}{3}\cm{3}{4}{2}{1\over\bk{34}[43]}={[12]^4\over[12][23][34][41]},
\ee
which is the Parke-Taylor result. 

Like the situation under item 1, gauge constraints of the vanishing $b_{ij}$ forbid us to apply $\d_i'$.  
Shift invariance under $\d_2, \d_3$ is trivial because $W_{[14]}=b_{14}$ is not shifted by any of them.

This calculation uses momentum conservation to combine the $\s$-factors into the standard form, after
which rules developed for double-color scalar amplitudes can be used to integrate. For $n>5$, it
is no longer possible to convert the $\s$-dependence into the standard form this way. Although there
are other methods to do so, they are fairly complicated and in any case do not seem to work well with 1-cycles.
In those cases it may be easier to do the integrations without combining the $\s$-factors. Let us illustrate how
that goes with the present example.

\subsubsection{\boldmath $1^+_{\l-}2^+3^-_+4^-_{\n}$}
The only non-zero $b$ is $b_{24}^{+-}$, thus the only allowed permutations are [124](3) and [1324].
With $+=3$ and $-=1$, $c_{13}=c_{23}=0=c_{31}=c_{41}$. Therefore
\be
\h\pf'\Q&=&-{W_{[124]}\over\s_{(124)}}C_{33}+{W_{[1324]}\over\s_{(1324)}}=-{W_{[124]}\over\s_{(124)}}C_{33},\nn\\
C_{33}&=&-{c_{32}\over\s_{32}}-{c_{34}\over\s_{34}}=c_{34}\({1\over\s_{32}}-{1\over\s_{34}}\)=c_{34}{\s_{24}\over
\s_{32}\s_{34}},\nn\\
\h\pf'\Q&=&{c^+_{12}b_{24}^{+-}c^-_{34}\over\s_{(1234)}},\nn\\
M&=&c^+_{12}b_{24}^{+-}c^-_{34}\({1\over k_2\.k_3}+{1\over k_1\.k_2}\)=-c^+_{12}b_{24}^{+-}c^-_{34}
{k_1\.k_3\over(k_1\.k_2)(k_2\.k_3)}\nn\\
&=&-\cp{1}{2}{3}\bb{2}{4}{3}{1}\cm{3}{4}{1}{\bk{13}[31]\over\bk{34}[43]\bk{23}[32]}={[12]^4\over[12][23][34][41]}.\ee
Momentum conservation identities $\bk{34}[41]+\bk{32}[21]=0$ has been used to
get the final result. Shift invariance is trivial under $\d_2$ and $\d_3$ because none of the factors of $\pf'\Q$ get shifted.
Shifts under $\d_i'$ should not be applied because of the gauge constraits.

Note that although the different choices of $(-,+)$ in \S VIIIA.1, 2, 3 all give the same final result,
they give rise to different cycle structures and different permutations, and also different $\s$-structure
before momentum conservation was used.

\subsection{\boldmath $1^+_{\l -}2^+_\n3^-4^-_+$}
Since the only non-vanishing $b$ is still $b_{23}^{+-}$, the allowed permutations are [1432] and [132](4). With $4=+$,
$c^+_{14}=c^+_{24}=0$. With $1=-$, $c_{31}=c_{41}=0$. Hence
\be
\h\pf'\Q&=&{W_{[1432]}\over\s_{(1432)}}-{W_{[132]}\over\s_{(132)}}C_{44}=b_{23}^{+-}\({c^+_{14}c^-_{43}\over\s_{(1432)}}-{c^+_{13}\over\s_{(132)}}C_{44}\)=-{b_{23}^{+-}c^+_{13}\over\s_{(132)}}C_{44},\nn\\
C_{44}&=&-{c_{41}^-\over\s_{41}}-{c_{42}^-\over\s_{42}}-{c_{43}^-\over\s_{43}}=-{c_{42}^-\over\s_{42}}-{c_{43}^-\over\s_{43}}
=c_{42}^-\({1\over\s_{43}}-{1\over\s_{42}}\)=c_{42}^-{\s_{32}\over\s_{43}\s_{42}},\nn\\
\h\pf'\Q&=&{b_{23}^{+-}c_{13}^+c_{42}^-\over\s_{(1243)}},\nn\\
M&=&-{2�b_{23}^{+-}c_{13}^+c_{42}^-\over s_{34}}=-\bb{2}{3}{4}{1}\cp{1}{3}{4}\cm{4}{2}{1}{1\over\bk{34}[43]}\nn\\
&=&-{\bk{34}[12]^2\over\bk{14}[41][43]}={[12]^4\over[12][23][34][41]}.\labels{4B}
\ee

\subsection{\boldmath $1^+_{\l -}2^+3^-_\n4^-_+$}
The only non-zero $b$ is $b_{23}^{+-}$, so the only allowed permutations are [123](4) and [1423]. Since $4=+$
and $1=-$, $c_{14}=c_{24}=0=c_{31}=c_{41}$. 
Thus
\be
\h\pf'\Q&=&-{W_{[123]}\over\s_{(123)}}C_{44}+{W_{[1423]}\over\s_{(1423)}},\nn\\
C_{44}&=&-{c_{42}\over\s_{42}}-{c_{43}\over\s_{43}}=c_{42}\({1\over\s_{43}}-{1\over\s_{42}}\)=c_{42}{\s_{32}\over\s_{42}\s_{43}},\nn\\
W_{[123]}&=&c^{+}_{12}b^{+-}_{23},\nn\\
W_{[1423]}&=&c^+_{14}c^-_{42}b_{23}^{+-}=0,\quad\Rw\nn\\
\h\pf'\Q&=&-{c^+_{12}b^{+-}_{23}c^-_{42}\over\s_{(1243)}}={c^+_{13}b^{+-}_{23}c^-_{42}\over\s_{(1243)}},
\ee
which is the same as \eq{4B}.

\subsection{\boldmath $1^+_{\l }2^+_-3^-_{\n+}4^-$}
In this case $c_{32}=c_{42}=0=c_{13}=c_{23}$, and the only non-zero $b$'s are $b_{14}$.
The allowed permutations are $A=[143](2)$. 
Choose the constant lines to be
1,3,4 so that the variable line is 2. The amplitude is then
\be M&=&{-1\over 2\pi i}\oint_\Gamma{2d\s_2 \over f_2}{\s_{(134)}^2b_{14}c_{34}C_{22}\over \s_{(1234)}\s_{(143)}}
=-{2\over s_{21}}b_{14}c_{34}c_{21}
\nn\\
&=&-{2\over s_{34}}\bb{1}{4}{3}{2}\cm{3}{4}{2}\cp{2}{1}{3}=-{\bk{43}[12]^2\over\bk{23}[43][32]}
={[12]^4\over[12][23][34][41]}.
\ee

\subsection{Summary for \boldmath$n=4$}
Table 2 presents a summary of the situations studied above. In spinor helicity language, they all reduce to the Parke-Taylor formula.

$$\ba{|c|c|c|c|}\hline
&{\tt configuration}&{\tt permutation}&M=[12]^4/[12][23][34][41]\\ \hline
A1&1^+_{\l-}2^+3^-4^-_{\n+}&[1234],[1324]&-b_{23}c_{12}c_{43}/k_3\.k_4\\
A2&1^+_\l2^+_-3^-_+4^-_\n&[14](2)(3)&-b_{14}c_{24}c_{34}/k_3\.k_4\\
A3&1^+_{\l-}2^+3^-_+4^-_\n&[1324],[124](3)&b_{24}c_{12}c_{34}(1/k_2\.k_3+1/k_3\.k_4)\\
B&1^+_{\l-}2^+_\n3^-4^-_+&[1432],[132](4)&-b_{23}c_{14}c_{42}/k_3\.k_4\\
C&1^+_{\l-}2^+3^-_\n4^-_+&[1423],[123](4)&-b_{23}c_{14}c_{42}/k_3\.k_4\\
D&1^+_{\l}2^+_-3^-_{\n+}4^-&[143](2)&-b_{14}c_{34}c_{21}/k_3\.k_4\\
\hline\ea$$
\bc Table 2.\quad Summary of the $n=4$ amplitude $M=M^{(1234)}$\ec

\section{\boldmath $n=5$: $1^+_{\lambda} 2^+_-3^{-}_{\n+}4^-5^-$}
In this case $c_{32}=c_{42}=c_{52}=0=c_{13}=c_{23}$, and the only non-zero $b$'s are $b_{14}$ and $b_{15}$.
Therefore the non-zero permutations contributing to $\pf'\Q$ are $A=[143](2)(5), B=[1453](2), C=[153](2)(4), D=[1543](2)$. 
Choosing the constant lines to be
1,3,5 and the variable lines to be 2 and 4 yields the amplitude 
\be M&=&\({-1\over 2\pi i}\)^2\oint_\Gamma{d\s_2 d\s_4\over f_2f_4}{\s_{(135)}^2\pf'\Q\over \s_{(12345)}},\nn\\
\pf'\Q&=&2^2C_{22}\({\a\over\s_{(143)}}C_{55}+{\b\over\s_{(1453)}}+{\g\over\s_{(153)}}C_{44}+{\d\over\s_{(1543)}}\),\nn\\
\a&=&b_{14}c_{34},\quad \b=b_{14}c_{54}c_{35},\quad \g=b_{15}c_{35},\quad \d=b_{15}c_{45}c_{34}.\labels{M5}
\ee

All four terms in $\pf'\Q$ have a positive sign. Actually, these plus signs are all of the form $(-1)^2$. To start with,
there is an overall minus sign in $\pf'\Q$. The permutation signature of the first term is $+$, but the surviving
matrix element in $\e_1U_4\e_3$ has a minus sign. The permutation signature of the second term is $-$, but the surviving
matrix element in $\e_1U_4U_5\e_3$ has a $+$ sign. The third term is like the first term and the fourth term is like
the second term. In this way, all the minus signs pile up to give a plus sign for all four terms.

Let us check the shift invariance of \eq{M5}. Since $\l=1$ and $\n=3$, we should check it only for $\d_2, \d_4, \d_5$. 
The invariance under
$\d_2$  is trivial because none of $\a,\b,\g,\d$ get shifted.
\eq{M5} also remains unchanged under $\d_4$ and $\d_5$, because the only non-zero $b$'s are $b_{14}$ and $b_{15}$,
and neither $c_{14}$ nor $c_{15}$ appears in $\a, \b, \g$, or $\d$. The shifts $\d'_i$ should not be applied because
of the gauge constraints.

We shall evaluate this integral in two ways. In subsection A, momentum conservation is used to combine the multi-cycle
$\s$-dependences into single 5-cycles, thus allowing rules for double-color scalar amplitudes to be applied to carry out the integrations.
This method will not work for $n>5$, in which case it is better to carry out direct integrations with the multi-cycle
structure present. We shall illustrate how this can be done in subsection B. Finally in subsection C, these results will be expressed
in spinor-helicity language to show that they agree with the Parke-Taylor formula.

\subsection{Conversion of multi-cycles into  single 5-cycles}
Using $c_{23}=0=c_{42}=c_{52}$ and the momentum conservation relation \eq{mcn}, we can write
\be
C_{22}&=&-{c_{21}\over\s_{21}}-{c_{24}\over\s_{24}}-{c_{25}\over\s_{25}}=c_{21}{\s_{51}\over\s_{25}\s_{21}}+c_{24}{\s_{54}\over\s_{25}\s_{24}}
=c_{21}{\s_{41}\over\s_{24}\s_{21}}+c_{25}{\s_{45}\over\s_{24}\s_{25}},\nn\\
C_{44}&=&-{c_{41}\over\s_{41}}-{c_{43}\over\s_{43}}-{c_{45}\over\s_{45}}=c_{41}{\s_{51}\over\s_{45}\s_{41}}+c_{43}{\s_{53}\over\s_{45}\s_{43}},\nn\\
C_{55}&=&-{c_{51}\over\s_{51}}-{c_{53}\over\s_{53}}-{c_{54}\over\s_{54}}=c_{51}{\s_{41}\over\s_{54}\s_{51}}+c_{53}{\s_{43}\over\s_{54}\s_{53}}.\labels{CCC}\ee
These relations allow us to write
\be
{C_{22}C_{55}\over\s_{(143)}}&=&+{c_{51}c_{21}\over\s_{(12543)}}-{c_{51}c_{24}\over\s_{(15243)}}-{c_{53}c_{21}\over\s_{(12453)}}+{c_{53}c_{25}\over\s_{(14253)}},\nn\\
{C_{22}\over\s_{(1453)}}&=&\hskip4cm {c_{21}\over\s_{(12453)}}-{c_{25}\over\s_{(14253)}},\nn\\
{C_{22}C_{44}\over\s_{(153)}}&=&-{c_{43}c_{21}\over\s_{(12543)}}+{c_{43}c_{24}\over\s_{(15243)}}+{c_{41}c_{21}\over\s_{(12453)}}-{c_{41}c_{25}\over\s_{(14253)}},\nn\\
{C_{22}\over\s_{(1543)}}&=&+{c_{21}\over\s_{(12543)}}-{c_{24}\over\s_{(15243)}}.\labels{CCC2}
\ee
Now that all the $\s$-dependences are of the double-color scalar type, we can use the trivalent \cite{CHY3}, polygon \cite{BBBD2}, or pairing
\cite{LYscalar} rule to evaluate the integral to get

\be
{1\over 4}M^\a&=&{1\over s_{12}}\({1\over s_{34}}+{1\over s_{45}}\)\(\a c_{51}c_{21}-\g c_{43}c_{21}+\d c_{21}\)
-{1\over s_{15}s_{43}}\(-\a c_{51}c_{24}+\g c_{43}c_{24}-\d c_{24}\)\nn\\
&&-{1\over s_{12}s_{45}}\(-\a c_{53}c_{21}+\b c_{21}+\g c_{41}c_{21}\)\nn\\
&=&{c_{21}\over s_{12}s_{45}}\(-\a c_{54}+\g c_{45}-\b+\d\)+
{c_{21}\over s_{12}s_{34}}\(\a c_{51}-\g c_{43}+\d \)\nn\\
&&-{c_{24}\over s_{15}s_{43}}\(-\a c_{51}+\g c_{43}-\d \).\labels{Mconvert}
\ee

This method to convert multi-cycles into a single 5-cycle no longer works for $n>5$ for the following reason. Each $C_{ii}$ consists
of $(n\-1)$ terms of the form $c_{ij}/\s_{ij}$, but with $c_{ik}=0$ where $k$ is the reference momentum of the appropriate helicity, there
remains only $(n\-2)$ terms. For $n=5$, one of the three terms is eliminated using momentum
conservation. If $c_{i\ell}$ is eliminated, then the factor $\s_{\ell j}\ (j\not=i,k,\ell)$ will appear in the numerator, as shown in \eq{CCC}.
To get the single 5-cycle structure shown in \eq{CCC2}. $\ell$ must be chosen so that every $\s_{\ell j}$ get cancelled out. Since 
in any cycle $\ell$ has only two nearest neighbors, this can be achieved with two $j$'s, as shown in \eq{CCC2}, but not with
three or more $j$'s, which is the case for $n>5$. For that reason this method of converting multi-cycle structures into single
a $n$-cycle structure cannot work for $n>5$. In that case we have to do the integrations without such a conversion. How this
can be done for $n=5$ is illustrated in the next subsection.

\subsection{Direct integrations}
As discussed in \S VI, to evaluate \eq{M5} directly, we must first determine the allowed crystal sets. These sets must contain
consecutive lines, including one constant line. With the constant lines
chosen to be 1,3,5, these sets are $S=\{21\}, \{23\}, \{234\}, \{43\}, \{45\}$. There are three terms resulting from the $\s_2$-integration,
coming from the first three $S$. The first two sets involve only $(2)=C_{22}=-c_{21}/\s_{21}-c_{23}/\s_{23}-\cdots$, but since
$c_{23}=0$, the pole $1/\s_{23}$ is absent, so that leaves only sets $\{21\}$ and $\{234\}$. If we denote the contribution of these two sets
to $M^\a$ to be respectively $M'$ and $M''$, then
\be {1\over 4}M'&=&{-1\over 2\pi i}\oint_{\Gamma_4}{ c_{21}d\s_4\over s_{21}f_4'}{\s_{(135)}^2\over \s_{(1345)}}
\({\a\over\s_{(143)}}C_{55}+{\b\over\s_{(1453)}}+{\g\over\s_{(153)}}C_{44}+{\d\over\s_{(1543)}}\),\nn\\
f_4'&=&{s_{41}+s_{42}\over\s_{41}}+{s_{43}\over\s_{43}}+{s_{45}\over\s_{45}}.\ee
To carry out the $\s_4$-integration, move the contour $\Gamma_4$ away from $f_4'=0$ to enclose the poles at $\s_{43}=0$
and $\s_{45}=0$. In this way we get
\be
{1\over 4}M'&=&{c_{21}\s_{(135)}\over s_{21}}\[\a\({c_{54}\over s_{45}\s_{(153)}}-{C_{55}'\over s_{34}\s_{(13)}}\)+{\b\over s_{45}\s_{(153)}}
+{\g\over\s_{(153)}}\({c_{43}\over s_{34}}-{c_{45}\over s_{45}}\)\right.\nn\\
&&\left. -{\d\over\s_{(153)}}\({1\over s_{34}}+{1\over s_{45}}\)\]\nn\\
&=&-{c_{21}\over s_{21}}\[\a\({c_{54}\over s_{45}}-{C_{55}'\s_{(153)}\over s_{34}\s_{(13)}}\)+{\b\over s_{45}}
+\g\({c_{43}\over s_{34}}-{c_{45}\over s_{45}}\) -\d\({1\over s_{34}}+{1\over s_{45}}\)\]\nn\\
&=&-{c_{21}\over s_{21}}\[\a\({c_{54}\over s_{45}}-{c_{51}\over s_{34}}\)+{\b\over s_{45}}
+\g\({c_{43}\over s_{34}}-{c_{45}\over s_{45}}\) -\d\({1\over s_{34}}+{1\over s_{45}}\)\]\nn\\
&=&{c_{21}\over s_{21}s_{34}}\(\a c_{51}-\g c_{43}+\d\)-{c_{21}\over s_{21}s_{45}}\(\a c_{54}+\b-\g c_{45}-\d\),\labels{Mp}
\ee
where $C'_{55}$ is $C_{55}$ evaluated at $\sigma_{43}=0$, which gives
$c_{51}{\sigma_{13}/ \sigma_{15}\sigma_{53}}$.

To evaluate  the contribution of the set $\{234\}$ to $M^\a$,  make a scaling change
$\s_{23}=s\s'_{23}=s, \s_{34}=s\s'_{34}$, and look for poles at $s=0$. As $s\to 0$,
\be
f_2&\to&{1\over s}\(s_{23}+{s_{24}\over\s'_{24}}\):={1\over s}f_2'',\nn\\
f_4&\to&{1\over s}\({s_{42}\over\s'_{42}}+{s_{43}\over\s'_{43}}\):={1\over s}f_4'',\nn\\
\s_{(12345)}&\to&s\s_{(135)}\s'_{34},\nn\\
C_{22}&\to&-{c_{24}\over s\s'_{24}},\nn\\
C_{55}&\to&-\({c_{51}\over\s_{51}}+{c_{52}+c_{53}+c_{54}\over\s_{53}}\)=c_{51}{\s_{31}\over\s_{51}\s_{53}},\nn\\
C_{44}&\to&-{1\over s}{c_{43}\over\s'_{43}}.\ee
After doing the $s$-integration, we obtain the contribution of $\{234\}$ to the amplitude to be
\be {1\over 4}M''
&=&{-1\over 2\pi i}\oint_{\Gamma_4}{ c_{24}d\s'_4\s_{(135)}\over \s'_{24}\s'_{34}f_2''f_4''}\({-\a c_{51}\over\s_{(13)}\s'_{43}}{\s_{31}\over\s_{51}\s_{53}}+{\g c_{43}\over\s_{(153)}\s'_{43}}+{\d\over\s_{(153)}\s'_{43}}\)\nn\\
&=&{-1\over 2\pi i}\oint_{\Gamma_4}{ c_{24}d\s'_4\over \s'_{24}{\s'_{34}}^2f_2''f_4''}\(-\a c_{51}
+\g c_{43}+\d\).\ee
The contour $\Gamma_4$ surrounds $f_4''=0$ counter clockwise. The pole for $f_4''=0$ is located at $\s'_{43}=s_{43}/(s_{42}+s_{43})$.
This means $\s'_{42}=\s'_{43}+\s'_{32}=-s_{42}/(s_{42}+s_{43})$, $f_2''=s_{23}+s_{34}+s_{24}=s_{51}$. Hence
\be
M''&=&4{c_{24}\over s_{34}s_{51}}\(\a c_{51}-\g c_{43}+\d\).\labels{Mpp}\ee
Combining \eq{Mp} and \eq{Mpp}, we get finaly
\be
M^\a=M'+M''&=&{4c_{21}\over s_{21}s_{34}}\(\a c_{51}-\g c_{43}+\d\)+{4c_{21}\over s_{21}s_{45}}\(-\a c_{54}-\b+\g c_{45}+\d\)\nn\\
&+&
{4c_{24}\over s_{34}s_{51}}\(\a c_{51}-\g c_{43}+\d\),\labels{Mfin}\ee
which is identical to the result obtained in \eq{Mconvert} by multi-cycle conversion.

\subsection{Spinor helicity expression}
Recall from \eq{M5} that
\be\a&=&b_{14}c_{34},\quad \b=b_{14}c_{54}c_{35},\quad \g=b_{15}c_{35},\quad \d=b_{15}c_{45}c_{34}.\nn\ee 
The three terms in \eq{Mfin} can then be written as
\be
X&=&{4c_{21}\over s_{21}s_{34}}\(\a c_{51}\-\g c_{43}\+\d\)={4c_{21}\over s_{21}s_{34}}\(b_{14}c_{34}c_{51}\+b_{15}(-c_{35}c_{43}
\+c_{45}c_{34})\),\nn\\
Y&=&{4c_{21}\over s_{21}s_{45}}\(-\a c_{54}\-\b+\g c_{45}\+\d\)={4c_{21}\over s_{21}s_{45}}\(b_{14}(-c_{34}c_{54}\-c_{54}c_{35})
\+b_{15}(c_{35}c_{45}\+c_{45}c_{34})\)\nn\\
&=&{4c_{21}\over s_{21}s_{45}}(-b_{14}c_{54}+b_{15}c_{45})(c_{34}+c_{35}),\nn\\
Z&=&{4c_{24}\over s_{34}s_{51}}\(\a c_{51}\-\g c_{43}\+\d\)={4c_{24}\over s_{34}s_{51}}\(b_{14}c_{34}c_{51}+b_{15}(-c_{35}c_{43}+c_{45}c_{34})\).\ee
The spinor-helicity expression for these quantities are
\be
c_{34}+c_{35}&=&\cm{3}{4}{2}+\cm{3}{5}{2}=-{\bk{13}[12]\over[32]},\nn\\
c_{35}c_{43}-c_{45}c_{34}&=&\cm{3}{5}{2}\cm{4}{3}{2}-\cm{4}{5}{2}\cm{3}{4}{2}={\bk{34}[52]\over[32][42]}\(\bk{53}[32]+\bk{54}[42]\)\nn\\
&=&{\bk{34}[52]\bk{15}[12]\over[32][42]},\nn\\
X&=&{4c_{21}\over s_{21}s_{34}}\(\bb{1}{4}{3}{2}\cm{3}{4}{2}\cm{5}{1}{2}-\bb{1}{5}{3}{2}{\bk{34}[52]\bk{15}[12]\over[32][42]}\)\nn\\
&=&{4c_{21}\over s_{21}s_{34}}{\bk{43}\bk{15}[12]^3\over[32][42][52]}={4\over s_{21}s_{34}}\cp{2}{1}{3}{\bk{43}\bk{15}[12]^3\over[32][42][52]}
=-{\bk{13}\bk{15}[12]^3\over\bk{23}\bk{21}[32][42][52][43]},\nn\\
Y&=&{4c_{21}\over s_{21}s_{45}}\(\bb{1}{4}{3}{2}\cm{5}{4}{2}-\bb{1}{5}{3}{2}\cm{4}{5}{2}\){\bk{13}[12]\over[32]}\nn\\
&=&-{4c_{21}\over s_{21}s_{45}}{\bk{13}\bk{45}[12]^3\over[32][42][52]}=-{4\over s_{21}s_{45}}\cp{2}{1}{3}{\bk{13}\bk{45}[12]^3\over[32][42][52]}	
=-{\bk{13}^2[12]^3\over\bk{23}\bk{21}[32][42][52][54]},\nn\\
Z&=&{4c_{24}\over s_{34}s_{51}}\(\bb{1}{4}{3}{2}\cm{3}{4}{2}\cm{5}{1}{2}-\bb{1}{5}{3}{2}{\bk{34}[52]\bk{15}[12]\over[32][42]}\)\nn\\
&=&-{4c_{24}\over s_{34}s_{51}}{\bk{15}\bk{43}[12]^3\over[32][42][52]}=-{4\over s_{34}s_{51}}\cp{2}{4}{3}{\bk{15}\bk{43}[12]^3\over[32][42][52]}
=-{\bk{43}[12]^3\over\bk{23}[32][52][43][15]},\nn\\
X+Y&=&-{\bk{13}[12]^3\over\bk{23}\bk{21}[32][42][52]}\({\bk{15}\over[43]}+{\bk{13}\over[54]}\)={\bk{13}[12]^3\over\bk{23}[32][43][52][54]}.\ee
putting all these together, we finally have
\be M^\a&=&X+Y+Z={[12]^3\over\bk{23}[32][43][52]}\({\bk{13}\over[54]}+{\bk{43}\over[51]}\)={[12]^4\over[12][23][34][45][51]},\ee
which agrees with the Parke-Taylor formula.

\section{Crystal Graph, $\s$-Table, and $C$-Table}
In \S IX B, integrations of a five-point amplitude were carried out directly, without first using algebraic manipulation
to bring its $\s$-dependence into the form of a double-color scalar amplitude. In this section, 
we discuss how to generalize that to an $n$-point amplitude, using the method of \S VI to get the denominator
and the numerator factors of the Feynman diagrams. The numerator is essentially $\pf'\Q$, after replacing the $\s_{ij}$ 
and the $C_{ii}$ in it by some appropriate values. The value for $\s_{ij}$ is
either $\pm 1, \pm\s_{rs}, \pm\s_{st}$, or $\pm\s_{tr}$,  
depending on what the Feynman diagram is, and how the integration is done. The precise value can be read off from
a `crystal graph' and  tabulated in a `$\s$-table'. How $C_{ii}$ turns out to be is
tabulated in a `$C$-table'.

The basis of the replacement stems from the following observations.
Let $S_1$ and $S_2$ be two non-intersecting crystals, with defects $r_1$ and $r_2$,
triggers $p_1$ and $p_2$, and integration variables $s_1$ and $s_2$, respectively. Denote the lines in $S_1\bsl\{r_1,p_1\}$
 by $a_1,b_1, \cdots,$ and $S_2\bsl\{r_2,p_2\}$ by $a_2,b_2,\cdots$, and the new $\s$-variables after the scaling change 
 in $S_1$ and $S_2$ by $\s'$ and $\s''$.
Then since the integral is evaluated in the vicinity of $s_1=0$ and $s_2=0$, the $\s_{ij}$ variables after the integrations turn into
\be
\s_{a_1b_1}&\to&\s'_{a_1b_1},\quad \s_{a_2b_2}\to\s''_{a_2b_2},\quad \s_{a_1a_2}\to\s_{r_1r_2},\nn\\
\s'_{p_1r_1}&=&1,\quad \s''_{p_2r_2}=1,\nn\\
\s_{Aa_1}&\to&\s_{Ar_1},\quad\s_{Aa_2}\to\s_{Ar_2},\quad (A\notin S_1, S_2).\labels{sr1}\ee
If $S_2\subset S_1$, then let $a_2,b_2,\cdots$ denote lines in $S_2$ as before, but let $a_1,b_1,\cdots$
denote lines in $S_1\bsl S_2$. Since $s_1$ is integrated before $s_2$, after the integrations, we have
\be
\s_{a_1b_1}&\to&\s'_{a_1b_1},\quad \s_{a_2b_2}\to\s''_{a_2b_2},\quad \s_{a_1a_2}\to\s'_{r_1a_2},\nn\\
\s'_{p_1r_1}&=&1,\quad \s''_{p_2r_2}=1,\nn\\
\s_{Aa_1}&\to&\s_{Ar_1},\quad\s_{Aa_2}\to\s_{Ar_1},\quad (A\notin S_1, S_2).\labels{sr2}\ee
Note that the defect $r_2$ for the smaller crystal $S_2$ must be either $r_1$ or $p_1$.

\subsection{Crystal graph and $\s$-table}
From the largest crystals we can proceed to create smaller and smaller crystals, using as a defect 
either a \M constant line, or a previous trigger. Each integration converts a variable line into a trigger line,
until every one of  the $(n\-3)$ variable lines has been turned into a trigger. 
The information of how integrations are carried out for a Feynman diagram, and the resulting relations using \eq{sr1} and \eq{sr2},
 can be summarized in a {\it crystal graph}.
Fig.~1 shows the Feynman diagram of an 8-point gauge amplitude and its crystal graph, with the constant lines indicated dashed 
in the Feynman diagram, and underlined in the crystal graph.
\bc\igw{3}{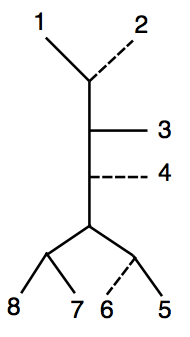}\qquad\igw{7}{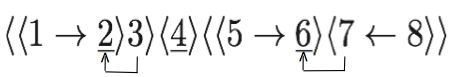}\\
Fig.~1.\quad An 8-point Feynman diagram and its crystal graph\ec

If we forget about the arrows,  then the crystal graph is simply an algebraic way to represent the Feynman diagram \cite{LYscalar}, with 
external lines forming a propagator grouped inside an angular bracket $\bk{\cdots}$. The arrows indicate the triggering relations $\s_{pr}=1$,
pointing from the trigger $p$ to the defect $r$. The manner integrations are carried out can be read out from the graph. In the case of Fig.~1, it tells
us that starting from two non-intersecting crystal $A=\{1\underline 23\}$ and $B=\{5\underline 678\}$, with defects $r=2$
and $r=6$ respectively, an integration is carried out in each crystal with respective triggers $p=3$ and $p=7$. After that, the triggers may
be used as defects for smaller crystals to be formed. Subsequent integrations are carried out in the crystal $\{12\}\subset A$,
with $2$ as the defect and $1$ as the trigger, and in the crystals $\{56\}\subset B$ and $\{78\}\subset B$, with 6 and 7 as the 
respective defects and 5 and 8 as the respective triggers.

Using the rules of \eq{sr1} and \eq{sr2}, the final expressions for $\s_{ij}$ can be read out from the crystal graph. For Fig.~1,
we have $1=\s_{12}=\s_{32}=\s_{56}=\s_{87}=\s_{76}$ because each pair is connected by an arrow. 
For the remaining $\s_{ij}$, move $i$ and $j$ along the arrows,
until the two are connected directly by an arrow, or both end up at some constant lines. Then read off the final $\s$ value at the end. 
If there are two conflicting ways of doing so, then the movement in the smaller crystal takes precedence. This is so because
 in a multi-crystal structure, as the scaling variables $s$ become small, the $\s_{ij}$
 within a smaller crystal is much smaller than that in the larger crystal. In this way the $\s$-table of Fig.~1 is obtained and shown in Table 3.

$$\ba{|c|cccccccc|}\hline
i\bsl j&1&2&3&4&5&6&7&8\\ \hline
1&0&+1&-1&\s_{24}&\s_{26}&\s_{26}&\s_{26}&\s_{26}\\
2&-1&0&-1&\s_{24}&\s_{26}&\s_{26}&\s_{26}&\s_{26}\\
3&+1&+1&0&\s_{24}&\s_{26}&\s_{26}&\s_{26}&\s_{26}\\
4&-\s_{24}&-\s_{24}&-\s_{24}&0&\s_{46}&\s_{46}&\s_{46}&\s_{46}\\
5&-\s_{26}&-\s_{26}&-\s_{26}&-\s_{46}&0&+1&-1&-1\\
6&-\s_{26}&-\s_{26}&-\s_{26}&-\s_{46}&-1&0&-1&-1\\
7&-\s_{26}&-\s_{26}&-\s_{26}&-\s_{46}&+1&+1&0&-1\\
8&-\s_{26}&-\s_{26}&-\s_{26}&-\s_{46}&+1&+1&+1&0\\
\hline\ea $$
 \bc Table 3.\quad The $\s$-table  for Fig.~1\ec
 
\subsection{$C$-table} 
 The $C$-table can be similarly obtained. For reasons mentioned at the end of last paragraph,
 the $j$-summation in $C_{ii}=-\sum_{j\not=i}c_{ij}/\s_{ij}$ can be truncated 
 into a sum over the smallest crystal $S_i$ containing $i$. Moreover, if $S_i$ contains an even smaller crystal $S_x$ with a 
 defect $r_x$, then all $\s_{ij}$ for $j\in S_x \subset S_i$ can be replaced by $\s_{ir_x}$. 
 After this replacement and truncation, the remaining $\s_{ij}$ in the sum should be replaced by those in the $\s$-table to
 get the $C$-table.
 
 For example, the $C$-table for the eight-point amplitude is
 $$\ba{|c|c|}\hline
 -C_{11}&c_{12}\\
 -C_{22}&-c_{21}\\
 -C_{33}&c_{31}+c_{32}\\
 -C_{44}&(c_{41}+c_{42}+c_{43})/\s_{42}+\\
 &(c_{45}+c_{46}+c_{47}+c_{48})/\s_{46}\\
 -C_{55}&c_{56}\\
 -C_{66}&-c_{65}\\
 -C_{77}&-c_{78}\\
 -C_{88}&c_{87}\\
 \hline\ea$$
 \bc Table 4.\quad The $C$-table for Fig.~1\ec

 Using \eq{pfprime}, the gauge amplitude in \eq{mg} is  given  by the formula
 \be
 M^{(123\cdots n)}=-2^{n-3}\s_{(rst)}^2\sum\(\pf'\Q\)\(\prod_S{1\over s_S}\),\labels{mint}\ee
 where the product is taken over a compatible set of $(n\-3)$ crystals $S$, and the sum is over all compatible crystal sets. 
 Each compatible set gives rise to  a crystal graph, a $\s$-table, a $C$-table, and 
 a product of $(n\-3)$ propagators $1/s_S$ that fixes the Feynman diagram involved.
The numerator of that Feynman diagram is obtained by substituting the relations of $\s_{ij}$ and $C_{ii}$
contained in the two tables into the expression for $\pf'\Q$. To be consistent, the $\s_{(rst)}^2$ factor must be cancelled
by the same factor contained in $\pf'\Q$.
 
 \subsection{Five-point amplitude}
 As shown in \eq{Mconvert}, the five-point amplitude computed in \S IX consists of three Feynman diagrams. We will illustrate the
 method outlined above by using it to compute the term with the propagator $1/s_{12}s_{45}$ , taking as before the constant
 lines to be 1, 3, 5.
 
 The Feynman diagram for this term and its crystal graph are shown in Fig.~2, from which
 we can obtain its $\s$-table shown in Table 5 and its $C$-table shown in Table 6. 
 \bc\igw{3}{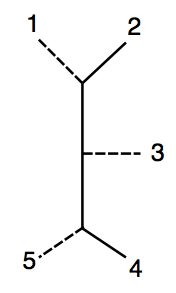}\qquad \igw{4.5 }{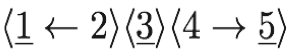}\\
 Fig.~2.\quad A Feynman diagram and its crystal graph for a five-point amplitude\ec
 
 $$\ba{|c|ccccc|}\hline
i\bsl j&1&2&3&4&5\\ \hline 
1&0&-1&\s_{13}&\s_{15}&\s_{15}\\
2&+1&0&\s_{13}&\s_{15}&\s_{15}\\
3&-\s_{13}&-\s_{13}&0&\s_{35}&\s_{35}\\
4&-\s_{15}&-\s_{15}&-\s_{35}&0&+1\\
5&-\s_{15}&-\s_{15}&-\s_{35}&-1&0\\
\hline\ea$$
\bc Table 5.\quad The $\s$-table for the amplitude in Fig.~2\ec

$$\ba{|c|c|}\hline
-C_{11}&-c_{12}\\
-C_{22}&c_{21}\\
-C_{33}&(c_{31}+c_{32})/\s_{31}+(c_{34}+c_{35})/\s_{35}\\
-C_{44}&c_{45}\\
-C_{55}&-c_{54}\\
\hline\ea$$
\bc Table 6.\quad The $C$-table for the amplitude in Fig.~2\ec

According to \eq{M5}, the reduced Pfaffian of the amplitude is given by
\be
\pf'\Q&=&2^2C_{22}\({\a\over\s_{(143)}}C_{55}+{\b\over\s_{(1453)}}+{\g\over\s_{(153)}}C_{44}+{\d\over\s_{(1543)}}\),\nn\\
\a&=&b_{14}c_{34},\quad \b=b_{14}c_{54}c_{35},\quad \g=b_{15}c_{35},\quad \d=b_{15}c_{45}c_{34}.\nn
\ee
To get the part of the amplitude proportional to $1/s_{12}s_{45}$, all we need is to use Fig.~2, or equivalently Tables 5 and 6,
to substitute in the expressions for $\s_{ij}$ and $C_{ii}$ in $\pf'\Q$. Then
\be
\pf'\Q&=&-4c_{21}\({-\a\over\s_{(135)} }c_{54}-{\b\over\s_{(135)}}+{\g\over\s_{(135)}}c_{45}+{\d\over\s_{(135)}}\).
\ee
Thus, according to \eq{mint}, this part of the amplitude is
\be 4c_{21}\(-\a c_{54}-\b_\g c_{45}+\d\),\ee
agreeing with the result \eq{Mconvert} obtained previously.

\section{CHY, Feynman Amplitudes, and the Four-Gluon Vertex}
In the usual field theory approach, 
a color-stripped Feynman tree amplitude is given by  a sum of  Feynman diagrams, constructed by using propagators to
link up triple-gluon and four-gluon vertices in all possible ways, while keeping the cyclic order of external lines fixed. 
In a covariant gauge, the numerator of each Feynman diagram consists of sums of products of the form
$p\.p', \e\.p, \e\.\e'$, where $\e, \e'$ are the polarization vectors of the external lines,
and $p, p'$ are internal or external momenta. To get the final result for each Feynman diagram, we must use momentum
conservation to express all internal momenta as sums of external momenta, then assemble and simplify the terms. The 
algebra involved is very complicated, the result varies from diagram to diagram, generally without any discernible  pattern.

In contrast, the CHY amplitude has a very regular numerator no matter what $n$ is. It contains $(n-1)!$ terms given by \eq{pfprime},
each with a distinct $\s$-structure. 
The corresponding denominators are obtained by carrying out integrations in $\s$. Neither
the triple nor the four gluon vertex appears explicitly. 

A Feynman amplitude is built up from local vertices. It has a complicated numerator but simple denominators in each Feynman
diagram. A CHY amplitude has a regular numerator and relatively complicated denominators that must be obtained by integration.
It is global; local structures can be extracted only after the amplitude is computed. 

In that connection, one puzzling feature about the CHY gauge amplitude is that it is closely related to the CHY scalar amplitude, 
which corresponds to a $\f^3$ coupling in field theory. Where is the four-gluon vertex coming from? Since the CHY
gauge amplitude is gauge invariant, if the triple-gluon vertex is contained in the CHY amplitude, which we know that it does, then
the four-gluon vertex must be present. To find out what it is we must carry out a computation of the $n=4$ CHY amplitude,
with general polarization vectors. Recall from \S VII that our normalization for $n=3$ corresponds to a coupling constant of $g=\h$
for the triple-gluon vertex.  
It turns out that in order to compare with the Feynman amplitude, the appropriate normalization for $n=4$ turns out to be $\h$ of that 
used in \eq{pfprime}.
In the rest of this section, we will carry out that calculation to find the four-gluon vertex from the CHY amplitude.


The $n=4$ amplitude is given by \eq{mg} to be
\be
M&=&-{1\over 2\pi i}\oint_\Gamma{d\s_4\over  f_4}{\s_{(123)}^2\pf'\Q\over\s_{(1234)}},\nn\\
 f_4&=&{  s_{41}\over\s_{41}}+{ s_{42}\over\s_{42}}+{  s_{43}\over\s_{43}}, \labels{me4general1}\ee
where the contour $\Gamma$ encircles $f_4=0$ counter-clockwise. 

The solution of $f_4=0$ is $\s_4=\s_4^0$, with
\be \s_4^0&=&{1\over x}\( s_{41}\s_2\s_{13}+ s_{41}\s_2\s_{23}\),\quad x= s_{41}\s_{13}+ s_{42}\s_{23}.\labels{sol1}\ee
This implies 
\be
\s_{41}^0={1\over x}\( s_{41}\s_{31}\s_{12}\),\quad \s_{42}^0={1\over x}\( s_{42}\s_{12}\s_{23}\),\quad 
\s_{43}^0={1\over x}\( s_{43}\s_{23}\s_{31}\),\labels{sol2}\ee
and
\be {1\over  f_4}={\s_{41}\s_{42}\s_{43}\over x(\s_4-\s_4^0)}.\ee
The amplitude $M$ is therefore given by the residue at $\s_4=\s_4^0$ to be
\be
M=L(\pf'\Q)_0,\quad L=-\({\s_{41}\s_{42}\s_{43}\s_{(123)}^2\over x\s_{(1234)}}\)_0={ s_{42}\s_{(123)}^2\over x^2},\ee
where the subscript 0 indicates that $\s_4$ should be evaluated at $\s_4^0$ given in \eq{sol1} and \eq{sol2}.

The reduced Pfaffian $\pf'\Q$ contains $3!=6$ terms with the following cycle structure:
 \be a=[142](3),\  b=[132](4),\ c=[12](3)(4),\ d=[12](34),\ e=[1342],\ f=[1432].\ee
 The reduced Pfaffian is given by \eq{pfprime} to be $-2$ times the sum of the following six quantities,
\be
(-)^a\Q_a&=&{\e_1U_4\e_2\over\s_{(142)}}C_{33},\nn\\
(-)^b\Q_b&=&{\e_1U_3\e_2\over\s_{(132)}}C_{44},\nn\\
(-)^c\Q_c&=&-{\e_1\.\e_2\over\s_{(12)}}C_{33}C_{44},\nn\\
(-)^d\Q_d&=&{\e_1\.\e_2\over\s_{(12)}}{\h\Tr(U_3U_4)\over\s_{(34)}},\nn\\
(-)^e\Q_e&=&-{\e_1U_3U_4\e_2\over\s_{(1342)}},\labels{psi4ng}\\
(-)^f\Q_f&=&-{\e_1U_4U_3\e_2\over\s_{(1432)}},\nn\\
\(C_{33}\)_0&=&\(-{c_{31}\over\s_{31}}-{c_{32}\over\s_{32}}-{c_{34}\over\s_{34}}\)_0=
{c_{32} s_{43}-c_{34} s_{41}\over s_{43}}{\s_{12}\over\s_{13}\s_{23}},\nn\\ 
\(C_{44}\)_0&=&\(-{c_{41}\over\s_{41}}-{c_{42}\over\s_{42}}-{c_{43}\over\s_{43}}\)_0=
{\( s_{41}\s_{13}+ s_{42}\s_{23}\)\( s_{41}\s_{12}- s_{43}\s_{23}\)\( s_{43}c_{41}- s_{41}c_{43}\)\over s_{41} s_{42} s_{43}\s_{12}\s_{13}\s_{23}}.\nn
\ee
Their contributions to $M$, denoted by $M_i=-2n_im_i, i=a,b,c,d,e,f$, are listed in  Table 7. The first column under $m_i$ gives the coefficients of $1/ s_{43}$, the second column gives the coefficients of $1/ s_{41}$, the third column displays other contributions. 

$$\ba{|c|c|c|c|c|c|}\hline
i&\multicolumn{2}{c|}{n_i}&\multicolumn{3}{c|}{m_i}\\ \cline{4-6}
&\multicolumn{2}{|c|}{}& s_{43}^{-1}& s_{41}^{-1}&{\rm others}\\ \hline\hline
a&\e_1U_4\e_2&-b_{14}c_{24}+c_{14}b_{24}&c_{34}&-c_{32}&0\\
b&\e_1U_3\e_2&-b_{13}c_{23}+c_{13}b_{23}&-c_{43}&c_{41}&0\\
c&\e_1\.\e_2&b_{12}&-c_{32}c_{43}-c_{34}c_{41}&c_{32}c_{41}&c_{34}c_{43}  s_{41}/  s_{43}^2\\
d&\e_1\.\e_2\h\Tr(U_3U_4)&b_{12}(c_{34}c_{43}-a_{34}b_{43})&0&0& s_{42}/  s_{43}^2\\
e&\e_1U_3U_4\e_2&b_{13}c_{43}c_{24}+c_{13}c_{34}b_{24}&1&0&0\\
&&-b_{13}a_{34}b_{24}-c_{13}b_{34}c_{24}&&&\\
f&\e_1U_4U_3\e_2&b_{14}c_{34}c_{23}+c_{14}c_{43}b_{23}&-1&-1&0\\
&&-b_{14}a_{43}b_{23}-c_{14}b_{43}c_{23}&&&\\
\hline\ea$$
\bc Table 7.\quad The result of \eq{me4general1} given by $M=-2\sum_{i=a,b,c,d,e,f}n_im_i$\ec

Since this calculation is relatively complicated, it is useful to check it
by  checking its shift invariance under \eq{singt}. The reduced Pfaffian is proportional to the sum of the terms
in \eq{psi4ng}. Since $C_{ii}$ is not to be shifted, and the different line in \eq{psi4ng} have difference $\s$-dependence,
$\pf'\Q$ is shift invariant if and only if each of the factors $n_i$ in Table 7 is invariant under the shifts $\d_3,
 \d_4, \d'_3,$ and $\d'_4$. The computation in the equation below shows that this is indeed the case.
\be
\d_3n_a&=&0,\quad \d_4n_a=-b_{14}b_{24}+b_{14}b_{24}=0,\nn\\
\d_3n_b&=&-b_{13}b_{23}+b_{13}b_{23}=0,\quad \d_4n_b=0,\nn\\
\d_3n_c&=&0,\quad \d_4n_c=0,\nn\\
\d_3n_d&=&b_{12}(b_{34}b_{43}-b_{34}b_{43})=0,\quad \d_4n_d=b_{12}(b_{34}c_{43}-c_{43}b_{43})=0,\nn\\
\d_3n_e&=&b_{13}b_{43}c_{24}+b_{13}c_{34}b_{24}-b_{13}c_{34}b_{24}-b_{13}b_{34}c_{24}=0,\nn\\
\d_4n_e&=&b_{13}c_{43}b_{24}+c_{13}b_{34}b_{24}-b_{13}c_{43}b_{24}-c_{13}b_{34}b_{42}=0,\nn\\
\d_3n_f&=&b_{14}c_{34}b_{23}+c_{14}b_{43}b_{23}-b_{14}c_{34}b_{23}-c_{14}b_{43}b_{23}=0,\nn\\
\d_4n_f&=&b_{14}b_{34}c_{23}+b_{14}c_{43}b_{23}-b_{14}c_{43}b_{23}-b_{14}b_{43}c_{23}=0;\nn\\
\d'_3n_a&=&0,\quad \d'_4n_a=-c_{14}c_{24}+c_{14}c_{24}=0,\nn\\
\d'_3n_b&=&-c_{13}c_{23}+c_{13}c_{23}=0,\quad \d'_4n_b=0,\nn\\
\d'_3n_c&=&0,\quad \d'_4n_c=0,\nn\\
\d'_3n_d&=&b_{12}(a_{34}c_{43}-a_{34}c_{43})=0,\quad \d'_4n_d=b_{12}(c_{34}a_{43}-a_{34}c_{34})=0,\nn\\
\d'_3n_e&=&c_{13}c_{43}c_{24}+c_{13}a_{34}b_{24}-c_{13}a_{34}b_{24}-c_{13}c_{43}c_{24}=0,\nn\\
\d'_4n_e&=&b_{13}a_{43}c_{24}+c_{13}c_{34}c_{24}-b_{13}a_{34}c_{24}-c_{13}c_{34}c_{24}=0,\nn\\
\d'_3n_f&=&b_{14}a_{34}c_{23}+c_{14}c_{43}c_{23}-b_{14}a_{43}c_{23}-c_{14}c_{43}c_{23}=0,\nn\\
\d'_4n_f&=&c_{14}c_{34}c_{23}+c_{14}a_{43}b_{23}-c_{14}a_{43}b_{23}-c_{14}c_{34}c_{23}=0.\nn\\
\ee

The third column of $m_i$ is non-zero only for $i=c$ and $d$. The reason why these two rows are different from
the others can be seen from \eq{psi4ng}. We have evaluated the integral \eq{me4general1} at the $f_4=0$
pole, but we could have evaluated it in a different way, by distorting the contour $\Gamma$ away from $ f_4=0$ to surround
the $\s_{4i}=0$ poles. If we do so, then we can see that rows $c$ and $d$ are different from
the others because both $\Q_c$ and $\Q_d$ have a double pole at $\s_{43}=0$ while every other term contains
only simple poles. In $\Q_d$ this is so because $\s_{(34)}=-\s_{43}^2$, and in $\Q_c$, this is because $C_{33}$
and $C_{44}$ can each contribute a term proportional to $1/\s_{43}$. It is these double poles that contribute
to the expression shown in the third row of $m_i$. Direct evaluation of double poles will be discussed in the next section.

These strange terms never appeared in our previous calculations because they are both proportional to $b_{12}$. 
In our previous calculations, we took particles 1 and 2 to have  the same
helicity, so in the helicity gauge $b_{12}=0$. 

At first sight these strange terms seem to spell trouble because Feynman propagators, at least in the Feynman gauge,
could only gives rise to terms proportional to $1/ s_{43}=1/s_{12}$ and $1/ s_{41}=1/s_{23}$, but not $1/s_{43}^2$. However, on closer examination,
we see that all the troublesome terms get cancelled out:
\be (n_cm_c+n_dm_d)_{others}&=&{b_{12}\over  s_{43}^2}\[c_{34}c_{43} s_{41}+(c_{34}c_{43}-a_{34}b_{43}) s_{42}\]\nn\\
&=&-{b_{12}\over  s_{43}}\(c_{34}c_{43}+\h b_{43} s_{42}\).\labels{nmcd}\ee
Another way of saying this is that the contribution from the double poles effectively cancels out.

Since we already know the triple-gluon vertex from \eq{tgv}, we can extract the four-gluon vertex from Table 7 by subtracting out the
$s_{43}$- and $s_{41}$-channel diagrams with two triple-gluon vertices. In a covariant gauge whose propagator is parametrized by $\xi$,
the $s_{43}$-channel diagram is
\be
{\cal S}&=&g^2\(\e_1\.\e_2k_{1\m}-\e_1\.\e_2k_{2\m}-2\e_2\.k_1\e_{1\m}+2\e_1.k_2\e_{2\m}\)\(g^{\m\n}+\xi (k_3+k_4)^\m(k_3+k_4)^\n/s_{43}\){1\over  s_{43}}\nn\\
&&\(\e_3\.\e_4k_{3\n}-\e_3\.\e_4k_{4\n}-2\e_4\.k_3\e_{3\n}+2\e_3.k_4\e_{4\n}\)
:={4g^2\over s_{43}}{\cal S}_0={{\cal S}_0\over s_{43}},\labels{s0}\ee
independent of $\xi$, where
\be{\cal S}_0={1\over 4}(b_{12}, -b_{12},-2c_{21},2c_{12})\bm{a_{13}&a_{14}&c_{31}&c_{41}\cr a_{23}&a_{24}&
c_{32}&c_{42}&\cr c_{13}&c_{14}&
b_{13}&b_{14}\cr c_{23}&c_{24}&b_{23}&b_{24}\cr}\em
\bm{b_{34}\cr -b_{34}\cr-2c_{43}\cr 2c_{34}\cr },\labels{s1234} \em \ee
and the last equality in \eq{s0} assumes $g=\h$.
Table 8 shows the difference  between $M/2$ obtained from Table 7 and ${\cal S}_0/ s_{43}$, 
for terms 
proportional to $ s_{43}^{-1}=s_{12}^{-1}$. Location $(i|j)$ indicates that  the expression is proportional to the 
$(i,j)$ element of the square matrix in \eq{s1234}.

 $$\ba{|c|c|c|c|c|c|}\hline
 {\rm term}&\multicolumn{2}{c|}{-\sum_in_im_i \rm\ in\ Table\ 7}&\multicolumn{2}{c|}{{\cal S}_0/ s_{43}}&A-B\\ \cline{2-5}
  s_{43}^{-1}&{\rm location}&{\rm value}=A&{\rm location}&{\rm value}=B&\\ \hline\hline
 b_{12}&c,d&-c_{31}c_{43}+c_{34}c_{41}&(1,2|3,4)&-c_{31}c_{43}+c_{34}c_{41}&0\\
 &&+\h b_{34} s_{24}&&+{1\over 4}b_{34}( s_{24}- s_{23})&-{1\over 4}b_{34} s_{12}\\
 b_{13}&b,e&c_{21}c_{43}+\h b_{24} s_{34}&(3|3)&c_{21}c_{43}&\h b_{24} s_{34}\\
 b_{14}&a,f&-c_{34}c_{21}-\h b_{23} s_{34}&(3|4)&-c_{21}c_{34}&-\h b_{23} s_{34}\\
 b_{23}&b,f&-c_{12}c_{43}&(4|3)&-c_{12}c_{43}&0\\
 b_{24}&a,e&c_{12}c_{34}&(4|4)&c_{12}c_{34}&0\\
 b_{34}&e,f&c_{13}c_{24}-c_{14}c_{23}&(3,4|1,2)&c_{13}c_{24}-c_{14}c_{23}&0\\
 \hline\ea$$
 \bc Table 8.\quad Terms  proportional to $ s_{43}^{-1}$\ec

Similarly, the $ s_{41}$-channel diagram is
\be
{\cal T}&=&{4g^2\over  s_{41}}{\cal T}_0={{\cal T}_0\over s_{41}},\ee
where
\be{\cal T}_0={1\over 4}(b_{23}, -b_{23},-2c_{32},2c_{23})\bm{a_{24}&a_{21}&c_{42}&c_{12}\cr a_{34}&a_{31}&c_{43}&c_{13}&\cr c_{24}&c_{21}&
b_{24}&b_{21}\cr c_{34}&c_{31}&b_{34}&b_{31}\cr}\em
\bm{b_{41}\cr -b_{41}\cr -2c_{14}\cr 2c_{41}\cr}. \em \ee
Table 9 shows the difference  between $M/2$ obtained from Table 7 and ${\cal T}_0/ s_{41}$, for terms 
proportional to $ s_{41}^{-1}=s_{23}^{-1}$. 

 $$\ba{|c|c|c|c|c|c|}\hline
 {\rm term}&\multicolumn{2}{c|}{-\sum_in_im_i \rm\ in\ Table\ 7}&\multicolumn{2}{c|}{{\cal T}_0/ s_{41}}&A-B\\ \cline{2-5}
  s_{41}^{-1}&{\rm location}&{\rm value}=A&{\rm location}&{\rm value}=B&\\ \hline\hline
 b_{12}&c,d&-c_{32}c_{41}&(3|4)&-c_{32}c_{41}&0\\
 b_{13}&b,e&c_{23}c_{41}&(4|4)&c_{23}c_{41}&0\\
 b_{14}&a,f&-c_{24}c_{32}+c_{34}c_{23}&(3,4|1,2)&-c_{24}c_{32}+c_{34}c_{23}&0\\
 &&-\h b_{23} s_{34}&&-{1\over 4}b_{23}(s_{23}+2 s_{12})&{1\over 4}b_{23}s_{23}\\
 b_{23}&b,f&-c_{13}c_{41}+c_{14}c_{43}&(1,2|3,4)&-c_{13}c_{41}+c_{14}c_{43}&0\\
 b_{24}&a,e&c_{14}c_{32}&(3|3)&c_{14}c_{32}&0\\
 b_{34}&e,f&-c_{14}c_{23}&(4|3)&-c_{14}c_{23}&0\\
 \hline\ea$$
  \bc Table 9.\quad Terms  proportional to $ s_{41}^{-1}$\ec 
 
The total difference between the CHY amplitude and the sum of the two
triple-vertex diagrams is given by the sum of the last column of Table 8, divided by $ s_{34}$, and the last column
of Table 9, divided by $ s_{41}$. It is equal to
 \be
- {1\over 4}b_{12}b_{34}-{1\over 4}b_{23}b_{41} +{1\over 2}b_{13}b_{24}, \labels{diff}\ee
which is just the four-gluon vertex for the color-stripped amplitude \cite{FL},
 \be D_{\a\b\g\d}=g^2(-g_{\a\b}g_{\g\d}+2g_{\a\g}g_{\b\d}-g_{\a\g}g_{\b\d}),\ee
 with $\a,\b,\g,\d$ being the Lorentz indices for particles 1, 2, 3, 4 and $g=\h$.
 
 \section{Double Pole}
 As discussed in the last section, between equations \eq{psi4ng} and \eq{nmcd}, a double pole is present at $\s_{43}=0$ for
 the $n=4$ amplitude. This double pole was absent in Sec.~VIII where the $n=4$ amplitude was computed 
 in the helicity gauge, which illustrates that with a suitable choice
 of the gauge parameters $r, s, t, \l, \n, +, -$, and in suitable helicity configurations, it may also be absent for larger-$n$ amplitudes. 
 In that case the multi-crystal technique discussed in previous sections can be used to evaluate the gauge amplitude for any $n$.
 However, for large $n$, there are situations when its presence cannot be avoided. In that case we must find a way to calculate the residue of  those
 terms containing a double pole. 
 
 There is of course no difficulty in principle to compute the residue of a double pole. It is equal to the derivative of the rest of the integrand
 evaluated at the pole. Since the derivative must act on the reduced Pfaffian, and on the scattering functions, the computation
 is very tedious even for small $n$. 
 The question is whether a simpler way can be found to calculate the residue without differentiation.
 The rest of this section is devoted to a discussion of this point.  This is what we refer to as the
 `double pole problem'. It is not a problem that prevents us from getting the final result of the gauge amplitude, 
 because residues at double poles can always be computed. It is just a question of whether such computations
 can be simplified.
  
 We will show in \S XIIA that there is a simple way to calculate the $n=4$ double pole encountered in the last section.
  This method is generalized to any  $n$ in \S XIIB, provided the double pole comes from a 2-cycle 
 and the corresponding product of two
 1-cycles. The discussion in \S XIIC shows that the double pole problem of a $k$-cycle can be reduced to
 a double pole problem of a $(k\-1)$-cycle, thus all double pole problems can be solved by induction.

 \subsection{\boldmath$n=4$} 
 The amplitude  \eq{me4general1} in the last section was evaluated at the simple pole  $f_4=0$. As mentioned there, 
 if the contour is distorted to have the integral evaluated at the poles $\s_{4i}=0$ instead, then a double pole at $\s_{43}=0$ is encountered
 for the terms $\Q_c$ and $\Q_d$ of $\pf'\Q$. In what follows we will compute these two terms with the distorted contour,
 to illustrate how a double pole can be evaluated by changing it into a product of two simple poles, with the help of the scattering equation.
 
 These two terms are given by \eq{me4general1} and \eq{psi4ng} to be
 \be
M_{cd}&=&-{1\over 2\pi i}\oint_{\Gamma}{d\s_4\over  f_4}{\s_{(123)}^2\pf'\Q_{cd}\over\s_{(1234)}},\nn\\
\pf'\Q_{cd}&=&{\e_1\.\e_2\over\s_{12}^2}\(C_{33}C_{44}+{\h\Tr(U_3U_4)\over\s_{43}^2}\),\labels{mcd1}\ee
with
\be
\e_1\.\e_2&=&b_{12},\nn\\
C_{33}&=&-{c_{31}\over\s_{31}}-{c_{32}\over\s_{32}}-{c_{34}\over\s_{34}},\nn\\
C_{44}&=&-{c_{41}\over\s_{41}}-{c_{42}\over\s_{42}}-{c_{43}\over\s_{43}},\nn\\
\h\Tr(U_3U_4)&=&c_{34}c_{43}-b_{34}a_{43}.\labels{mcd2}\ee

The integrand consists of  simple poles at $\s_{41}=0$ and at $\s_{43}=0$, and a double pole at $\s_{43}=0$. The simple poles can
be evaluated in the usual way resulting in
\be
M_{cd}^{simple}={b_{12}\s_{(123)}\over\s_{12}^2}\[{c_{41}\over s_{41}}\({c_{31}+c_{34}\over\s_{31}}+{c_{32}\over\s_{32}}\)-
 \({c_{31}\over\s_{31}}+{c_{32}\over\s_{32}}\){c_{43}\over s_{43}}+\({c_{41}\over\s_{31}}+{c_{42}\over\s_{32}}\){c_{34}\over s_{43}} \]\ee
Using momentum conservation which implies $c_{31}=-c_{32}-c_{34}$ and $c_{42}=-c_{41}-c_{43}$, this can be reduced to
\be M_{cd}^{simple}=b_{12}\({c_{41}c_{32}\over s_{41}}- {c_{32}c_{43}+c_{41}c_{34}+c_{34}c_{43}\over s_{43}}\).\ee

The double-pole contribution to $\pf'\Q$ is
\be
{b_{12}\over\s_{12}^2}\[-c_{34}c_{43}+(c_{34}c_{43}-b_{34}a_{43})\]{1\over\s_{43}^2}=-\h {b_{12}\over\s_{12}^2}{b_{34}s_{43}\over\s_{43}^2}.\labels{noc3}\ee
Since it is proportional to $s_{43}/\s_{43}$, the double pole can be transformed into the product of two simple poles using the scattering equation $f_4=0$ to get
 \be
 \h{b_{12}\over\s_{12}^2}{b_{34}\over\s_{43}}\({s_{41}\over\s_{41}}+{s_{42}\over\s_{42}}\).\labels{dts}\ee
 Of course this has to be done before distorting the contour $\Gamma$ away from $f_4=0$. After this change, we can distort the contour and 
 evaluate the resulting simple poles at $\s_{43}=0$ and $\s_{41}=0$ in the usual way to get
 \be
 M_{cd}^{double}=\h{b_{12}b_{34}\s_{(123)}\over\s_{12}^2}\[-{1\over s_{43}}\({s_{41}\over\s_{31}}+{s_{42}\over\s_{32}}\)
 +{1\over \s_{13}}\]=-\h{b_{12}b_{34}s_{42}\over s_{43}},
\ee
where the last expression is obtained by using momentum conservation to replace $s_{41}$ with $-s_{42}-s_{43}$.

The final result
\be
M_{cd}=M_{cd}^{simple}+M_{cd}^{double}=b_{12}\[{c_{41}c_{32}\over s_{41}}- {1\over s_{43}}\(c_{32}c_{43}+c_{41}c_{34}+c_{34}c_{43}+\h b_{34}s_{42}\)\]
\ee
agrees with the result of rows $c$ and $d$ of Table 7, after taking into account equation \eq{nmcd}.

 \subsection{2-cycles}
 Assuming as usual that $\s_\a=\s_{(12\cdots n)}$ in \eq{mg}. Let $i,j$ be two of the $n$ external lines, then both the 2-cycle $(ij)$
 and the product of two 1-cycles $(i)(j)$ contain $1/\s_{ji}^2$. With this $\s_\a$, the factor $1/\s_{ij}^2$ yields a double pole in $s$
 only when $i$ and $j$ are adjacent, {\it viz}., when $j=i\pm 1$.  
 We will also assume that neither $i$ nor $j$ is at
 the two ends of an open cycle, a condition that can be satisfied by a choice of $\l$ and $\n$.
 
Consider any term in $\pf'\Q$ of the form $X(ij)$ and  $X(i)(j)$, where $X$ represents the rest of the cycle structure in $\pf'\Q$.
Then their contribution to $\pf'\Q$ is similar to that in \eq{mcd1} and \eq{mcd2}, and can be written as
\be
\Q_{ij}&=&\Q_X\(\Q_{(i)(j)}-\Q_{(ij)}\)=\Q_X\(C_{ii}C_{jj}+{c_{ij}c_{ji}-b_{ij}a_{ji}\over\s_{ij}^2}\),\nn\\
C_{ii}&=&{c_{ij}\over\s_{ij}}+\sum_{k\not=i,j}{c_{ik}\over\s_{ik}},\nn\\
C_{jj}&=&{c_{ji}\over\s_{ji}}+\sum_{k\not=i,j}{c_{jk}\over\s_{jk}}.
\ee
As in the last subsection, the contribution of $\pf'\Q_{ij}$ to \eq{mg} is made up of simple and double poles. Simple
poles are evaluated in the way discussed in earlier sections, so let us concentrate on the double pole. It is 
\be
\pf'\Q_{ij}={\Q_X\over\s_{ij}^2}\[-c_{ij}c_{ji}+(c_{ij}c_{ji}-b_{ij}a_{ji})\]=\h\Q_X{b_{ij}\over\s_{ij}}{s_{ij}\over\s_{ij}}.\labels{2cyc}\ee
Once again,  scattering equation can be used to replace $s_{ij}/\s_{ij}$ with $-\sum_{h\not=i,j} s_{ik}/\s_{ik}$, thereby
converting the double to a product of two simple poles. The rest can be evaluated as before, using a method similar to that
of Secs.~VI, X, and XIIA.

\subsection{$k$-cycles}
The crucial step that allows the double-pole problem in a 2-cycle to be solved is equation \eq{noc3}, in which
terms quadratic in $c$ disappear, leaving behind an expression proportional to $s_{43}/\s_{43}^2$. Then using
the scattering equation, the double-pole can be transformed into product of two simple poles as is done in \eq{dts}.
If the quadratic term in $c$ did not vanish, then of course the scattering equation can still be used to
replace one $1/\s_{43}=(s_{43}/\s_{43})/s_{43}$, but then we would end up with a term proportional to $1/s_{43}^2$
which could not be the whole story, as Feynman diagrams only allow a simple-pole propagator $1/s_{43}$ but not a
double pole term $1/s_{43}^2$.

For $k\ge 3$, it turns out that this crucial step still holds: terms of $\pf'\Q$ of degree-$k$ in $c$ add up to zero.
To see more explicitly how that happens, let us consider  
$k=3$, with the three consecutive lines in the 3-cycle to be 1,2,3. 
In the crystal $\{12\underline 3\}$ with $3$ being the defect,
double-pole appear in $\pf'\Q$ from the cycles (123), (132),
(12)(3), (31)(2), (23)(1), and (1)(2)(3). Together with the signature factor, their respective contributions to $\pf'\Q$ are
\be
\Q_{(123)}&=&{1\over 2\s_{12}\s_{23}\s_{31}}\(-c_{12}c_{23}c_{31}+c_{12}b_{23}a_{31}-c_{21}a_{23}b_{31}+c_{21}c_{32}c_{13}\)
\nn\\
&+&{1\over 2\s_{12}\s_{23}\s_{31}}\(-b_{12}c_{32}a_{31}+a_{12}c_{23}b_{31}+b_{12}a_{23}c_{31}-a_{12}b_{23}c_{13}\),\nn\\
\Q_{(132)}&=&{1\over 2\s_{13}\s_{32}\s_{21}}\(-c_{13}c_{32}c_{21}+c_{13}b_{32}a_{21}-c_{31}a_{32}b_{21}+c_{31}c_{23}c_{12}\)
\nn\\
&+&{1\over 2\s_{13}\s_{32}\s_{21}}\(-b_{13}c_{23}a_{21}+a_{13}c_{32}b_{21}+b_{13}a_{32}c_{21}-a_{13}b_{32}c_{12}\),\nn\\
-\Q_{(12)(3)}&=&-{c_{12}c_{21}-b_{12}a_{21}\over\s_{12}\s_{21}}C_{33},\nn\\
-\Q_{(31)(2)}&=&-{c_{31}c_{13}-b_{31}a_{13}\over\s_{31}\s_{13}}C_{22},\nn\\
-\Q_{(23)(1)}&=&-{c_{23}c_{32}-b_{23}a_{32}\over\s_{23}\s_{32}}C_{11},\nn\\
\Q_{(1)(2)(3)}&=&C_{11}C_{22}C_{33},
\labels{3cyc}\ee
with
\be
C_{11}&=&-{c_{12}\over\s_{12}}-{c_{13}\over\s_{13}}-\cdots,\nn\\
C_{22}&=&-{c_{21}\over\s_{21}}-{c_{23}\over\s_{23}}-\cdots,\nn\\
C_{33}&=&-{c_{31}\over\s_{31}}-{c_{32}\over\s_{32}}-\cdots.\ee
All these terms in $\pf'\Q$ scale like $1/s^3=1/s^{m+1}\ (m+1=k)$, which produces a double pole $1/s^2$ in the integrand of \eq{mg}.
The ellipses in \eq{CCC} represent terms that do not contribute a double pole so they will be ignored.

It can be verified by direct algebraic computation
that the $c^3$ terms in \eq{3cyc} add up to zero. As stated before, the $c^k$ terms in a $k$-cycle
also add up to zero for any $k$, and there is a simple reason for that. If we let 
$x_{ij}=c_{ij}/\s_{ij}$ for $i\not=j$, and $x_{ii}=C_{ii}$. The products of
$k$ $c$'s add up to zero  because the sum is proportional to the determinant of the $k\x k$ matrix $(x_{ij})$.
This determinant vanishes because its row sums are zero.

With the $c^3$ terms gone, a factor $s_{12}/\s_{12}, s_{23}/\s_{23}$,
or $s_{31}/\s_{31}$ is always present in every remaining term of $\pf'\Q$. If we could use the three scattering equations
$f_1=f_2=f_3=0$ to solve them in terms of $s_{ir}/\s_{ir}$, with $1\le i\le 3$ and $r>3$, then we would
have transformed every double pole in $\pf'\Q$ into a product of simple poles, and the problem is solved.

Unfortunately this cannot be done because the matrix to be inverted for the linear equation is singular, so
the three scattering equations allows only two $x_{ij}=s_{ij}/\s_{ij}\ (1\le i,j\le 3)$ to be solved, but not 3.
Moreover, the solution of these two depends on the value of the third, so the double pole cannot be
gotten rid of this way, for it still resides in the pole of the third quantity. For example, $x_{23}$ and $x_{13}$
can be solved in terms of $x_{i2}$ and $x_{ir}$ with $1\le i\le 3$ and $r>3$. As a result, the double pole involving
$s_{12}/\s_{12}^2$ is still present.

However, this remaining double pole involves only lines 1 and 2, and can be solved by the method of \S XIIB.
For example, consider the 8-point amplitude whose Feynman diagram and crystal graph are shown
in Fig.~1. The relevant part of the crystal graph is $\bk{\bk{1\rightarrow 2}3}$, with $p=3$ being the trigger
and $r=2$ being the defect of the larger crystal $\{123\}$. With this consideration, the double pole problem of the 
3-cycle (123) in the larger crystal becomes
the double pole problem in the smaller crystal $\{12\}$ for the 2-cycle (12).

Note that this procedure relies on the fact that the $\s_{ij}$ in a smaller crystal is much smaller than
that in a larger crystal. For that reason,  when $f_1=0$ is used to solve for $x_{12}$,
the term $x_{13}$ can be ignored. Otherwise the problem is not solved because the solutions
would have become circular.

Similarly, we know that in the presence of a $k$-cycle involving lines $\{123\cdots k\}$, 
the $c^k$
terms in $\pf'\Q$ together with those from all the other relevant product cycles add up to zero. 
We will use the letters $i,j$ to denote numbers between 1 and $k$, and the letter 
$r$ to denote a numbers larger than $k$.
In every remaining term in $\pf'\Q$, at least one factor
$x_{ij}=s_{ij}/\s_{ij}$ is present, and altogether there are $k(k-1)/2$ such factors. Using the the scattering equations
$f_i=0$, which is of rank $k-1$, we can solve $k-1$ of the $x_{ij}$ in terms of the rest. Let us choose those
$k-1$ factors to be $x_{ak}$, with $1\le a\le k-1$. After that, all the double pole terms become double pole terms
involving $x_{ab}$, where $1\le a,b\le k-1$, thereby reducing the double pole problem of the $k$-cycle into a
double pole problem for $(k\-1)$-cycle. Since $\s_{ij}$ in a smaller crystal is much smaller
than that in a larger crystal, by induction, all double pole problems can be solved this way.

\section{Summary}
The evaluation of the CHY $n$-point gauge tree amplitude in \eq{mg} is discussed. There are two basic difficulties: how
to handle the $(2n-3)!!$ terms in the reduced Pfaffian, and how to carry out the $(n\-3)$ integrations in the presence of a
$\s$ dependence much more complicated than the Parke-Taylor form found in the CHY double-color scalar amplitude.

We found a way to solve the first difficulty by grouping together  terms of the reduced Pfaffian into 
open and closed cycles of the  permutation group $S_n$.
We discovered a shift invariance for individual cycles which can be used to check  explicit calculations. 
The use of helicity gauge to further
simplify calculations has also been amply discussed. 

Integrations are carried out by dividing the dominant integration regions into $(n\-3)$ compatible `crystals',
each possessing a pole that can be evaluated in one of the $(n\-3)$ integrations using residue calculus. 
A complete recipe is provided when only simple poles are present to extract
 the $(n\-3)$ Feynman diagram propagators, with different compatible crystals giving rise  to different Feynman diagrams. 
A recipe making use of the `crystal graph' is also given to convert the $\s_{ij}$ factors in the reduced Pfaffian
into constants, after which the reduced Pfaffian can be identified as the numerator of the Feynman diagram.
Double poles are rarely present in small-$n$ amplitudes, but when they appear in large-$n$ amplitudes,
there will be more contribution to the amplitude in addition to what is given by the recipes above. These double poles 
can also be handled.

Calculations depend on a set of gauge parameters: the constant lines $r,s,t$, the Pfaffian lines $\l,\n$, and
the helicity reference lines $+,-$ in the helicity gauge. Although the final answer must not depend on the choice,
they have to be fixed in a calculation, and the structure and complexity of the calculation depend on their choice.
Many examples are given for the $n=3, n=4, n=5$ amplitudes to illustrate these calculations.

Unlike the Feynman amplitude, the CHY amplitude is global without the built-in local parts in the form
of vertices and propagators. Propagators turn out to emerge from the scattering functions $f_i$, but the vertices
are harder to extract. We have been successful in extracting the 
 triple-gluon vertex  from the $n=3$ amplitude, and the  four-gluon vertex 
from the $n=4$ amplitude. The calculation for the latter is somewhat lengthy. 

Both the CHY formula and the Feynman-diagram technique give the same scattering amplitude, but
they group their terms differently. Roughly speaking, the CHY amplitude groups by numerator,
and the Feynman amplitude groups by denominator. 
The numerator factors in  CHY amplitudes are universal for all $n$, 
being the cycles of the reduced Pfaffian. Its
denominators, however, have to be computed by integration, and a single numerator cycle may involve
propagators from several Feynman diagrams. In contrast, Feynman amplitudes are arranged according to
Feynman diagrams whose propagators are easy to write down, but its numerator factors have to be built up
from products of triple and four gluon vertices. Detailed algebraic manipulation which depends on the topology
of the Feynman diagram is needed to expand into its final form as a function of $\e_i\.\e_j, \e_i\.k_j,$ and $k_i\.k_j$.

This comparison of the two approaches is based on a general helicity configuration. For special
helicity configurations, it is often simpler to calculate in the helicity gauge, for either formalism.

Since complication of a gluon amplitude comes largely from its numerator factors, especially for large $n$,
there is a definite advantage in computing it in the numerator-grouping approach.
Another advantage is that the CHY formula has many
invariances that can be used to check the calculations. They all stem from the basic \M invariance
of the CHY amplitude, including the shift invariance, and the independence of the amplitude on the choice
of the gauge parameters $r, s, t, \s_r, \s_s, \s_t, \l, \n, +, -$. The disadvantage of the CHY formalism
is the lack of visible local structures in terms of propagators and vertices, to enable the underlying physical process to
be easily understood.

\end{document}